\documentclass[12pt]{article}

\usepackage{newtxtext,newtxmath}
\usepackage[T1]{fontenc}
\usepackage{siunitx}
\usepackage{graphicx}

\usepackage[letterpaper,margin=1in]{geometry}

\renewenvironment{abstract}
	{\quotation}
	{\endquotation}

\date{}

\makeatletter
\renewcommand{\fnum@figure}{\textbf{Figure \thefigure}}
\renewcommand{\fnum@table}{\textbf{Table \thetable}}
\makeatother

\usepackage{scicite}

\usepackage{url}

\def\scititle{Tunneling spectroscopy of topological insulator Josephson junctions}
\title{\bfseries \boldmath \scititle}

\author{
    Jakob~Schluck$^{1\dagger\ast}$,
	Ella~Nikodem$^{1\dagger}$,
	Anton~Montag$^{2,3,4}$,
	Alexander~Ziesen$^{2}$,\and
    Mahasweta~Bagchi$^{1}$,
    Fabian~Hassler$^{2}$,
    Yoichi~Ando$^{1\ast}$\and
	\small$^{1}$Physics Institute II, University of Cologne, Zülpicher Str. 77, 50937 Köln, Germany.\and
	\small$^{2}$Institute for Quantum Information, RWTH Aachen University, 52056 Aachen, Germany.\and
    \small$^{3}$Max Planck Institute for the Science of Light, 91058 Erlangen, Germany. \and
    \small$^{4}$Department of Physics, Friedrich-Alexander-Universit\"at Erlangen-N\"urnberg, 91058 Erlangen, Germany.\and
	\small$^\ast$Corresponding authors. Email: schluck@ph2.uni-koeln.de, ando@ph2.uni-koeln.de\and
	\small$^\dagger$These authors contributed equally to this work.
}

\begin{document} 

\maketitle

\begin{flushleft}
{\bf Short title:}\\
Topological insulator Josephson junction spectroscopy
\end{flushleft}

\begin{flushleft}
{\bf Teaser: }
The supgap spectrum of a topological surface state coupled to superconductors is studied experimentally.
\end{flushleft}

\begin{abstract} \bfseries \boldmath
In the seminal proposal by Fu and Kane, the superconducting proximity effect is used to realize topological superconductivity in the topological surface state (TSS) of a 3D topological insulator (TI).
In a Josephson junction made on the TI surface, the spin-momentum locking of the TSS guarantees the existence of spin-non-degenerate, perfectly transmitted Andreev modes. These modes lead to gap closing as a function of the superconducting phase difference $\varphi$.
Here, we report the observation of highly transmissive Andreev modes in a TI Josephson junction, probed via tunnel contacts and $\varphi$ controlled by a flux loop. Within our experimental resolution, they could indicate a gap closing and reopening. This phenomenon is observed for a wide range of chemical potentials, supporting its TSS origin. Under an applied magnetic field, highly transmissive modes are observed locally, matching theoretical predictions for Josephson vortices in the TSS. 
Our results support the fundamental role of spin-momentum locking in the Andreev physics in the TSS.
\end{abstract}

\noindent

\section{Introduction}

After more than 100 years from the discovery, superconductors remain a fascinating subject in physics. They become even more interesting when topology comes into play \cite{Sato2017}. A special type of superconductor, called ``spinless chiral $p$-wave superconductor'', is a prototypical topological superconductor hosting Majorana zero-modes at point-defects \cite{Kitaev2001, Sato2017}.
The Majorana zero-modes obey non-Abelian exchange statistics \cite{Read2000, Ivanov2001}, which make them a promising candidate for topological quantum computing \cite{Nayak2008}. Due to this exciting prospect for applications, the realization of topological superconductivity and the observation of Majorana zero-modes have become major themes in condensed matter physics.

Since a spinless chiral $p$-wave superconductor is not yet found in nature, there have been substantial efforts to effectively realize it in a hybrid system \cite{Alicea2012, Flensberg2021}. Theoretically, a hybrid of a strongly spin-orbit-coupled material and a conventional superconductor would effectively realize the spinless chiral $p$-wave pairing due to spin-momentum locking. The earliest of such proposals made by Fu and Kane in 2008 is based on the topological surface state (TSS) of a 3D topological insulator (TI) proximitized by a conventional $s$-wave superconductor \cite{Fu2008}. 
However, even though the spin-momentum locking is most naturally realized in TIs, the experimental evidence for topological superconductivity obtained so far in this platform is weak and relies on indirect signatures, such as modified Fraunhofer patterns \cite{Williams2012, Kurter2013, Yue2024} and the microwave response \cite{Wiedenmann2016, Deacon2017, Schuffelgen2019} of a Josephson junction made on a TI, or a broad zero-bias conductance peak (ZBCP) observed in the Abrikosov vortex core of a proximitized TI \cite{Xu2015}.
The search for Majorana zero-modes has been more advanced on  one-dimensional (1D) semiconducting Rashba nanowires, proximity-coupled to a superconductor, and exposed to a strong magnetic field \cite{Flensberg2021}; however, despite a remarkable experimental progress \cite{Krogstrup2014, Aghaee2023}, the distinction between topological and trivial ZBCPs remains contentious in this 1D platform \cite{Valentini2021, Dassarma2023, Hess2023}. 
It was recognized early on \cite{Akhmerov2011} and emphasized in the recent literature \cite{Rosdahl2018,Pikulin2021,Banerjee2023b,Banerjee2023c} that a direct observation of the closing and reopening of the superconducting (SC) gap, which implies the topological phase transition, is a more robust signature of topological superconductivity and thus Majorana zero-modes.

In this work, we turn to the TI platform and report the observation of highly transmissive Andreev modes in a TI Josephson junction, which could facilitate a gap closing and reopening upon changing the phase difference $\varphi$ across the junction. The spin-momentum locking of the TSS prohibits exact backscattering, and hence the normal reflection is forbidden for a perpendicular mode at the boundary to a superconductor. Such an absence of normal backscattering in turn guarantees perfect Andreev reflection, leading to a pair of perfectly transmitted, non-degenerate Andreev bound states (ABSs) in a short Josephson junction. 
These perfectly transmitted modes become gapless for $\varphi = \pi$, and the spin-non-degeneracy of the TSS makes the electron and hole branches to have opposite parity \cite{Tkachov2013}. As $\varphi$ is varied, these non-degenerate modes cross and the ground-state parity alternates at $\varphi = (2n+1)\pi$ with $n$ integer \cite{Fu2008}. Since the fermion parity is the $\mathbb{Z}_2$ topological invariant of a 1D superconductor \cite{Kitaev2001}, the gap closing and reopening corresponds to a topological phase transition. It is important to emphasize that the appearance of the perfectly transmitted ABSs is a direct consequence of the spin-momentum locking of the TSS that prohibits backscattering. Therefore, as long as the Josephson effect takes place in the TSS, the topological phase transition as a function of $\varphi$ is a necessity. Nevertheless, this fundamental property has not been experimentally confirmed so far in TI Josephson junctions.

Note that the Fu-Kane theory is only valid for a constant $\varphi$ along the junction. When an out-of-plane magnetic field is applied to the junction, rendering $\varphi$ position-dependent \cite{Potter2013}, the gapless 1D mode becomes confined at the core of a Josephson vortex, where the local phase difference is $\varphi = (2n+1)\pi$. Because of this, the gapless 1D mode evolves into a Majorana zero-mode that is bound to the point defect formed by the Josephson vortex. 

In our experiment, we observe strongly dispersing ABSs by tuning $\varphi$ across the junction in a SQUID geometry and simultaneously detecting the local density of states (LDOS) at different locations, indicating highly transmissive modes. They are observed over a wide range of chemical potentials and can be interpreted within our experimental resolution as a gap closing and reopening. In addition to the modes with the strongest dispersion, there is a multitude of low-lying excited states, as expected from theory. Importantly, our interpretation is not affected by whether a ZBCP is observed or there is just a smooth closing of the gap.
We further elucidate the locality of the involved states created by the phase gradient in perpendicular magnetic fields.
While there are alternative scenarios that could generate a similar phenomenology, our observations can most naturally be understood as consequences of the spin-momentum locking in the TSS. Based on this promising result, we suggest concrete conditions to be realized in future experiments to nail down the topological superconductivity and to generate well-protected Majorana zero-modes.

\section{Results}

\subsection{Phase-biased TI Josephson junction}

\begin{figure}
\centering
\includegraphics[width=\textwidth]{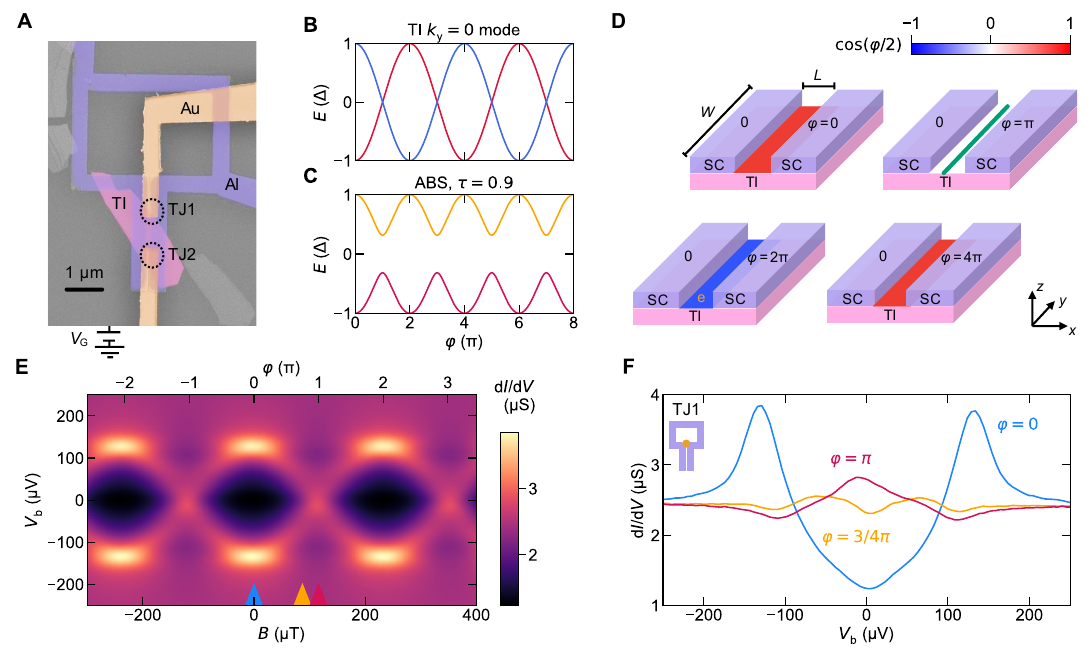}
\caption{\linespread{1.05}\selectfont{}
{\bf Phase-dependent Andreev bound states in a TI Josephson junction.} 
\textbf{A} False-color SEM image of a typical device. The exfoliated TI flake (pink) is contacted by Al  (lavender) forming an RF-SQUID geometry. Metallic contacts (Au) allow for probing the LDOS at two positions (TJ1 and TJ2) in the junction. \textbf{B,C,} Phase-dispersions of the perpendicular ($k_y$ = 0) mode in a TI line junction ({\bf B}) and conventional Andreev bound states (ABSs) with $\tau=0.9$ ({\bf C}). In panel {\bf B}, the red and blue lines correspond to the even- and odd-parity bound states, respectively. 
\textbf{D,} Evolution of the topological phase in the junction; here, the junction area is colored according to the gap function $m \propto \cos \frac{\varphi(y)}{2}$ (see the color scale). As $\varphi$ is tuned, the nature of the gap changes from trivial ($\varphi$ = 0, red) to topological ($\varphi$ = $2\pi$, blue). This is accompanied by a change in ground state parity from even to odd, as indicated by the extra electron (yellow) in the junction. In-between, the gap has to close ($\varphi=\pi$) via a gapless Majorana mode (green line). As the phase difference reaches $\varphi=4\pi$, a full period is completed. 
\textbf{E,} color mapping of d$I$/d$V$ in the $B$ vs $V_{\rm b}$ plane measured on TJ1 at 50 mK with $V_{\rm G}$ = 0 V. At $\varphi$ = 0 (mod $2\pi$), a proximity-induced gap opens with conductance peaks at $\pm$130 $\upmu$V. The gap closes and a broad ZBCP is observed at $\varphi = \pi$. A periodicity of 230 $\upmu$T is observed, consistent with the device geometry. 
\textbf{F,} Line-cuts of the data in {\bf E} at three selected magnetic fields corresponding to $\varphi$ = 0, $\frac{3}{4}\pi$, and $\pi$.}
\label{fig:JJ}
\end{figure}

The ABS spectrum in superconductor-normal-superconductor (SNS)-type Josephson junctions is commonly studied via tunneling spectroscopy by embedding the junction in a SQUID geometry \cite{leSueur2008,Pillet2010,Bretheau2017,Ren2018,Fornieri2018,Nichele2020}. In particular, Ren et al.\cite{Ren2018} and Fornieri et al.\cite{Fornieri2018} employed this technique to look for a topological phase transition in semiconducting Josephson junctions. They relied on the Zeeman effect due to an in-plane magnetic field to lift the spin-degeneracy of the ABS and detect the emerging ZBCPs. Here we apply this technique to the bulk-insulating TI platform, where the topological phase transition does not require a large magnetic field and can be tuned by the phase alone.
A scanning electron microscope (SEM) image of our typical device is shown in Fig.~\ref{fig:JJ}A. A Josephson junction is created by a thin film of Al on top of an exfoliated flake of the bulk-insulating TI material BiSbTeSe\textsubscript{2} \cite{Ren2011,Arakane2012}. The length $L$ of the junction is about $50~\textrm{nm}$, which is shorter than the surface state coherence length (see \ref{SI:sscl}), making it a short junction to host a single 1D Andreev mode \cite{Bretheau2017}.  The two sides of the junction are connected by an Al loop, which allows us to accurately tune the phase difference $\varphi$ by applying a perpendicular magnetic field. 
This device geometry does not allow us to measure the critical current $I_{\rm c}$, but our previous study of $I_{\rm c}$ in a similar TI Josephson junction \cite{Ghatak2018} confirmed that the supercurrent is primarily flowing through the TSS.

Normal electrodes made of Au are deposited after putting a thin layer of $\mathrm{Al_2O_3}$ as a tunnel barrier. Due to the native oxide of Al, its surface can be treated as an insulator and the overlap to the Au electrodes does not contribute to the transport. A back-gate voltage can be applied via the degenerately-doped silicon substrate. Therefore, this setup enables us to directly probe the LDOS of the TI Josephson junction and to test the Fu-Kane theory \cite{Fu2008}.

The perpendicular mode of the ABSs (i.e. zero longitudinal momentum $k_y$) appearing in a line junction made on the TSS is expected to present a $4\pi$-periodic $\varphi$-dependence as shown in Fig.~\ref{fig:JJ}B in the short-junction limit. As mentioned above, this is dictated by the spin-momentum locking which prohibits backscattering. The corresponding evolution of the gap in the junction is depicted in Fig.~\ref{fig:JJ}D: The gap is trivial for $0 \leq \varphi < \pi$. It closes at the topological phase transition with $\varphi=\pi$ where the gapless 1D mode (green line) appears, before reopening in the topological regime, with $\pi < \varphi < 3\pi$. In the topological regime the parity of the ground state is odd, meaning that the bound state below the chemical potential $\mu$ is filled with an unpaired electron (shown in yellow). Because of this, the junction returns to the initial state only at $\varphi = 4\pi$. In contrast, for a conventional Andreev bound state, a $2\pi$-periodicity is expected, as depicted in Fig.~\ref{fig:JJ}C. Here, the dispersion is given by \cite{Beenakker1991}
\begin{equation}
E=\pm\Delta\sqrt{1-\tau \sin^2\left(\varphi/2\right)}\,\,, 
\label{eq:ABS}
\end{equation}
with $\tau$ the transmittance of the state and $\Delta$ the induced SC gap in the normal metal  beneath the SC electrodes. In such a scenario, the gap can close in the case of perfect transmission ($\tau$ = 1), but the ground-state parity remains even across the gap closing, as conventional ABSs are doubly-degenerate.

Note that experimentally, when the perpendicular magnetic field is very weak such that the large loop is threaded by only a few flux quanta (and no Josephson vortex nucleates in the junction), the phase $\varphi$ is almost constant along the junction, a situation akin to the Fu-Kane model. The case at larger magnetic field, when Josephson vortices nucleate in the junction, is discussed later. 

We focus on the results obtained from a device similar to the one shown in Fig.~\ref{fig:JJ}A with two independent tunnel junctions TJ1 and TJ2 located at different positions on the same Josephson junction. Additional data obtained from a second, essentially identical device, is shown in the supplementary material \ref{SI:repro} to demonstrate the reproducibility of our results.  

\subsection{Tunnel spectra}
In Fig.~\ref{fig:JJ}E, we show the continuous evolution of the differential conductance d$I$/d$V$ of TJ1 with the applied DC bias voltage $V_{\rm b}$ and the perpendicular magnetic field $B$ around zero. Line cuts for selected fields are shown in Fig.~\ref{fig:JJ}F. Without magnetic field applied, we observe sharp particle-hole symmetric peaks at the bias voltages of $\pm 130\,\mathrm{\upmu eV}$ and a strong reduction in d$I$/d$V$ at lower bias.
The constant d$I$/d$V$ at higher bias corresponds to a resistance of about $400\, \mathrm{kohms}$, indicating we are in the tunneling regime, where the observed d$I$/d$V$ is proportional to the LDOS. 
The peak energy of $130\, \mathrm{\upmu eV}$ is smaller than the SC gap of our Al-film, which can be calculated from the critical temperature to be $170\, \mathrm{\upmu eV}$ (see \ref{SI:TC}). In a different device without TI but conditions that allowed for probing the Al gap directly, we found a density of states consistent with that of a typical BCS superconductor (see \ref{SI:BCS}). This confirms that the LDOS of the proximitized TI surface, not the LDOS of the Al electrode, is being measured. 
The LDOS at low bias does not drop to zero but remains finite, which indicates additional in-gap states that are not fully proximitized, consistent with our previous result on similar TI Josephson junctions, where the average transparency was $\tau_{\rm av} \approx0.8$ \cite{Ghatak2018}. Such a soft gap has also been reported for comparable devices based on carbon nanotubes \cite{Pillet2010}, graphene \cite{Bretheau2017},  HgTe \cite{Ren2018}, or InAs \cite{Fornieri2018}. Its most likely origin are channels with long quasi-classical trajectories that only couple weakly to the superconductors and can thus have low-lying energy levels with weak phase-tunability \cite{Bretheau2017,Ren2018}. Since for such channels the short junction approximation is no longer valid, their existence is not captured by models not considering the dynamical phase buildup as in Refs. \cite{Fu2008} and\cite{Tkachov2013}. 
However, it is well known that ballistic proximity systems, if they are extended enough, as well as diffusive ones, if they are disordered enough, will not present a hard induced gap\cite{Pilgram2000}. In the ballistic case, this effect can be attenuated by specifically designing the device to cut off long trajectories, as was proposed in Ref.\cite{Laeven2020}.

As a finite phase bias $\varphi$ is generated across the junction with the application of an out-of-plane magnetic field, the positions of the particle-hole symmetric peaks shift to a lower bias voltage and their amplitude is diminished. For the magnetic field of about $115\, \mathrm{\upmu T}$ that corresponds to $\varphi = \pi$, the particle-hole symmetric peaks merge into a single broad peak at zero bias, signaling the closing of the gap.  The spectrum then evolves periodically, with the conductance at $230\,\mathrm{\upmu T}$ becoming identical to the one observed at zero field. Having the Fu-Kane theory in mind, our data is consistent with the interpretation that the peak positions trace the gap, corresponding to the lowest-lying state at zero momentum, that evolves with $\varphi$ and closes at $\pi$ (see Fig.~\ref{fig:JJ}B).
In the brighter regions above this gap (but still below the bulk gap), there is a continuum of states corresponding to finite momentum $k_y$ along the junction \cite{Fu2008}, {which is indeed reflected in our LDOS spectra. Within this interpretation, the gap closing at $\varphi=\pi$ corresponds to a topological phase transition \cite{Fu2008}.} 
Upon close inspection, a certain degree of asymmetry in the LDOS between positive and negative energies can be found in the tunneling conductance. This is a typical phenomenon in similar devices \cite{Bretheau2017,Ren2018,Fornieri2018} and among the proposed mechanisms behind this effect are non-equilibrium effects\cite{Martin2014}, sub-gap states in the superconductor\cite{Stenger2017}, and voltage-dependent tunnel barriers\cite{Melo2021}. Also, the matrix elements describing the coupling between the normal probe and the ABS are unknown.
It is useful to note that similar devices made on various materials in the ballistic regime \cite{Pillet2010, Ren2018, Fornieri2018} with $\tau < 1$ exhibited gapped spectra at $\varphi = \pi$, consistent with Eq.~\eqref{eq:ABS}. The apparent gap closing observed here is most likely due to the unique property of the TSS, which necessitates the existence of a $\tau = 1$ mode.

\begin{figure}
\centering
\includegraphics[width=\textwidth]{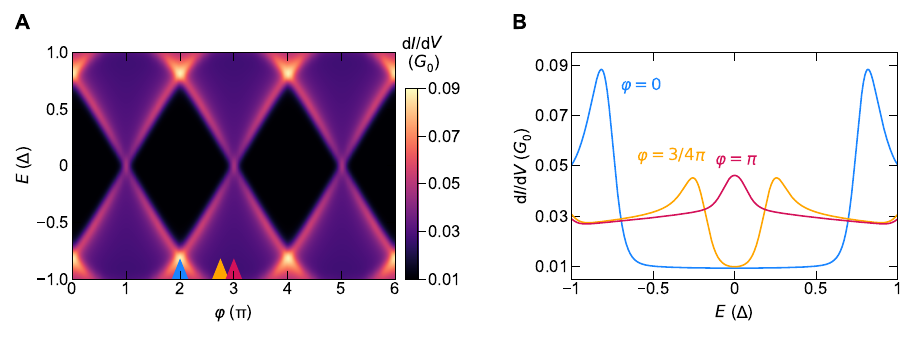}
\caption{\linespread{1.05}\selectfont{}
{\bf Theoretical simulations.} 
\textbf{A,} Color-mapping of the simulated differential conductance d$I$/d$V$ in units of the conductance quantum $G_0=e^2/h$ as a function of $V_{\rm b}$ and $\varphi$ for a temperature $T= \Delta/(30 k_{\rm B})$ in the Fu-Kane regime with $\varphi$ constant along the junction. 
\textbf{B,} Simulated behavior of the differential conductance d$I$/d$V$ at $\varphi$ = 0, $\frac{3}{4}\pi$, and $\pi$, which correspond to those shown in Fig.~\ref{fig:JJ}F. Note that in the simulation only the surface states and a single transversal mode are taken into account which explains the softer-gap in the experimental results of Fig.~\ref{fig:JJ}. 
}
\label{fig:Theory}
\end{figure}

To further corroborate this interpretation, we have performed numerical simulations of the differential conductance $G$  for a low-energy effective model of a TI Josephson junction (see methods section for details). Note that we do not aim to accurately model the experimental situation, but rather want to demonstrate how they agree with predictions from a rather minimal model.
Figure~\ref{fig:Theory}A shows the calculated $G$ as a function of $V_{\rm b}$ and $\varphi$, while the line cuts for the same $\varphi$ values as in Fig.~\ref{fig:JJ}F are shown in Fig.~\ref{fig:Theory}B. 
These simulations confirm that the even- and odd-parity bound states having a $4\pi$-periodicity produce the two branches that cross at $\varphi = (2n+1)\pi$, as observed in experiments. 
In Fig.~\ref{fig:Theory}B, the coherence peaks show up at the edge of the excitation gap and they diminish rapidly with increasing $\varphi$. 
Given the limited size of the simulated system, the bulk states above the gap are not correctly captured. Thus, we emphasize that the simulation is only accurate at low energies, below $0.5\Delta$, and that the position of the coherence peak cannot be traced to larger energies.
Our simulations suggest that the exact shape of the spectrum at $\varphi = \pi$ depends sensitively on the experimental conditions, and the gap closing can result in a nearly flat d$I$/d$V$ (see \ref{SI:simfix}). All these results are consistent with our experimental observations. It should be emphasized that the gap closing always occurs at $\varphi = (2n+1)\pi$, but it does not need to be accompanied by a ZBCP, according to our simulations. 

\subsection{Locality of the spectra in perpendicular magnetic fields}

\begin{figure}
\centering
\includegraphics[width=9cm]{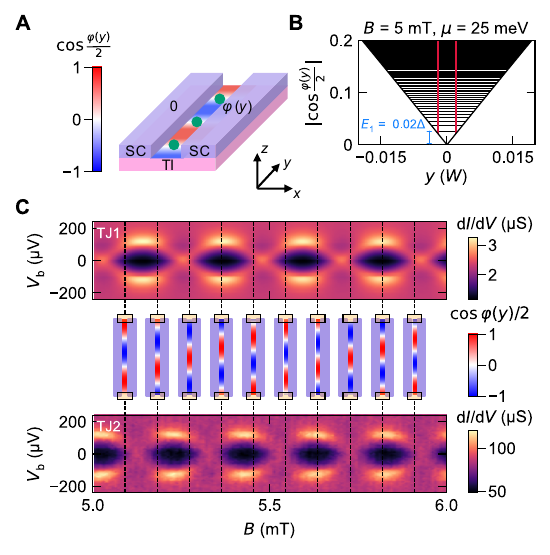}
\caption{\linespread{1.05}\selectfont{}
{\bf Locality of the spectra. } 
\textbf{A,} Schematic drawing of a planar TI Josephson junction in a perpendicular magnetic field, which generates a position-dependent phase $\varphi(y)$ across the junction. The distribution of the topological gap function $m \propto \cos \frac{\varphi(y)}{2}$ is depicted in the junction with a color scale. The position of $\varphi(y) = (2n+1)\pi$ corresponds to the Josephson vortex-core, where the gap closes (white color) and a Majorana zero-mode (green dot) appears.
\textbf{B,} Position-dependent gap $|m|=|\cos \frac{\varphi(y)}{2}|$ near a Josephson vortex-core in a magnetic field of $B$ = 5 mT, plotted with the excited states calculated from the model of Ref.~\citenum{Potter2013} for a chemical potential $\mu$ = 25 meV and $\Delta$ = 0.17 meV (thin horizontal lines), which form a quasi-continuum with a minigap of only 2\% of $\Delta$. The vertical red lines mark the spatial extent of the expected Majorana zero-mode appearing around $y$ = 0.
\textbf{C,} color mapping of d$I$/d$V$ in the $B$ vs $V_{\rm b}$ plane for TJ1 (top panel) and TJ2 (bottom panel) measured in the magnetic field range of 5 -- 6 mT at 50 mK with $V_{\rm G} = -5$ V. Note the slight difference in the periodicity, which is 230 $\upmu$T for TJ1 and  200 $\upmu$T for TJ2. The middle panel depicts the expected distribution of the topological gap function in a TI Josephson junction, based on our device geometry, calculated with Eq.~\eqref{eq:phase} evolves with magnetic fields; the position where the gap function locally becomes zero with $\varphi(y) = (2n+1)\pi$ corresponds to local gap closing (white regions). }
\label{fig:HighB}
\end{figure}

Next, we study the behavior of our device in higher magnetic fields and compare the LDOS probed at TJ1 and TJ2. When a perpendicular magnetic field $B$ is applied to the junction, $\varphi$ becomes dependent on the position along the junction ($y$ direction). This locality is reflected in the bound-state spectrum, as recently demonstrated in a semiconductor 2DEG-based junction \cite{Moehle2022}. If $B$ is strong enough to nucleate Josephson vortices, the position where the local phase difference becomes $\varphi = (2n+1)\pi$ is identified as the Josephson vortex-core, {and in TI Josephson junctions,} the gapless 1D mode in the Fu-Kane theory for global $\varphi = (2n+1)\pi$ turns into a Majorana zero-mode confined to the Josephson vortex-core \cite{Potter2013}. The concept of this Majorana zero-mode generation is sketched in Fig.~\ref{fig:HighB}A. A spatially varying phase $\varphi$ along a junction caused by $B$ leads to a position-dependent gap function $m\propto \cos \frac{\varphi(y)}{2}$ as indicated by the color code \cite{Potter2013}. Associated with the sign change in the gap function, the gap closes locally at $\varphi(y) = (2n+1)\pi$ (white area), corresponding to the core of the Josephson vortex, and a Majorana zero-mode (green dots) nucleates that is confined along the junction by the phase gradient. We note that the topologically-protected nature of the gap closing is intuitively understandable in this situation: The region of the junction with $-\pi < \varphi(y) < \pi$ (red) has the even-parity ground state with $\mathbb{Z}_2$ index of 0, whereas the region with $\pi < \varphi(y) < 3\pi$ (blue) has the odd-parity ground state with $\mathbb{Z}_2$ index of 1; therefore, the boundary between the two regions must become gapless for the topological invariant to change. 

We indeed observed the locality of the ABSs spectra, reacting to the local phase difference in a finite magnetic field. The top and bottom panels of Fig.~\ref{fig:HighB}C show the d$I$/d$V$ data simultaneously measured at TJ1 and TJ2 as a function of $V_{\rm b}$ and $B$ in the 5 -- 6 mT range.
Both junctions present a comparable evolution of their bound state spectrum, but there is a clear phase shift and their periods are slightly different. For TJ2 we find a periodicity of $200\, \mathrm{\upmu T}$, smaller than that of TJ1 ($230\, \mathrm{\upmu T}$). This difference in the period is consistent with the sample geometry (see \ref{SI:geom} for details) in which the effective junction area considering the flux focusing effect \cite{Ghatak2018, Moehle2022,Banerjee2023} is about $\sim$10\% of the loop area. 
To understand the LDOS at TJ1 and TJ2, we calculated the expected texture of the gap function of a TSS based on the device geometry.
In our junctions, the self-inductance is negligible and their width is much shorter than the Josephson penetration depth (see \ref{SI:JoPen} and \ref{SI:LI}), justifying the description of the phase gradient in terms of
\begin{equation}
\varphi(y)=\frac{2\pi B}{\Phi_0} \left (A + y L_{\rm eff}  \right)\,\, ,
\label{eq:phase}
\end{equation}
with $A$ the loop area, $\Phi_0=\frac{h}{2e}$ the magnetic flux quantum, and $L_{\rm eff}$ the effective length of the junction taking the flux-focusing effect into account. The evolution of the gap function $m\propto \cos\frac{\varphi(y)}{2}$ is calculated from this formula with the loop and junction area as fitting parameters (see \ref{SI:geom}) and visualized in the middle panel of Fig.~\ref{fig:HighB}C. We show snapshots of the phase-texture in the junction for a range of magnetic fields corresponding to the experimental situation. In this range, there are two to three Josephson vortices inside the junction, whose position is tuned by the global perpendicular magnetic field. They enter the junction from the bottom where TJ2 is located, and move upwards to TJ1 with increasing magnetic field. Whenever a Josephson vortex-core crosses the location of a tunnel junction, the energy gap appears to close. This is in line with the local nature of the gap closing predicted in the presence of Josephson vortices.  {Note that in a similar experiment on a non-topological Josephson junction \cite{Moehle2022}, the gap was found to remain open at the Josephson vortex-core.}
Our data shown in Fig.~\ref{fig:HighB}C was taken at an applied back-gate voltage of $V_{\rm G}=-5$~V, and we chose the magnetic field range of 5 -- 6~mT for the visibility of the phase gradient. The same pattern was found to continue into higher magnetic fields until the Al superconductivity was suppressed at 15 mT (see \ref{SI:HighB}).

We note that the spectra shown in Fig.~\ref{fig:HighB}C essentially look the same as those in Fig.~\ref{fig:JJ}E. This is because the Majorana zero-mode in the Josephson vortex-core is not well-separated in energy. For the conditions realized, it is accompanied by a quasi-continuum of low-energy bound states \cite{Potter2013}. Taking the smearing due to the finite-temperature into account, the local spectrum becomes a broad peak, indistinguishable from the Fu-Kane situation. 
In Fig.~\ref{fig:HighB}B, we show the magnitude of the gap function near a Josephson vortex-core and the finite-energy bound states calculated from the model in Ref.~\citenum{Potter2013}, along with the spatial extent of the expected Majorana zero-mode. Assuming a chemical potential of $\mu$ = 25 meV that is not very close to the Dirac point, as is the case in our device, the calculated finite-energy bound states form a quasi-continuum with a minigap of only 2\% of $\Delta$ (i.e. 3 $\upmu$eV), which is of the order of the thermal energy at the base temperature. For higher chemical potentials this energy separation is expected to shrink further.

\subsection{{Stability of the spectra against gating}}

\begin{figure}
\centering
\includegraphics[width=\textwidth]{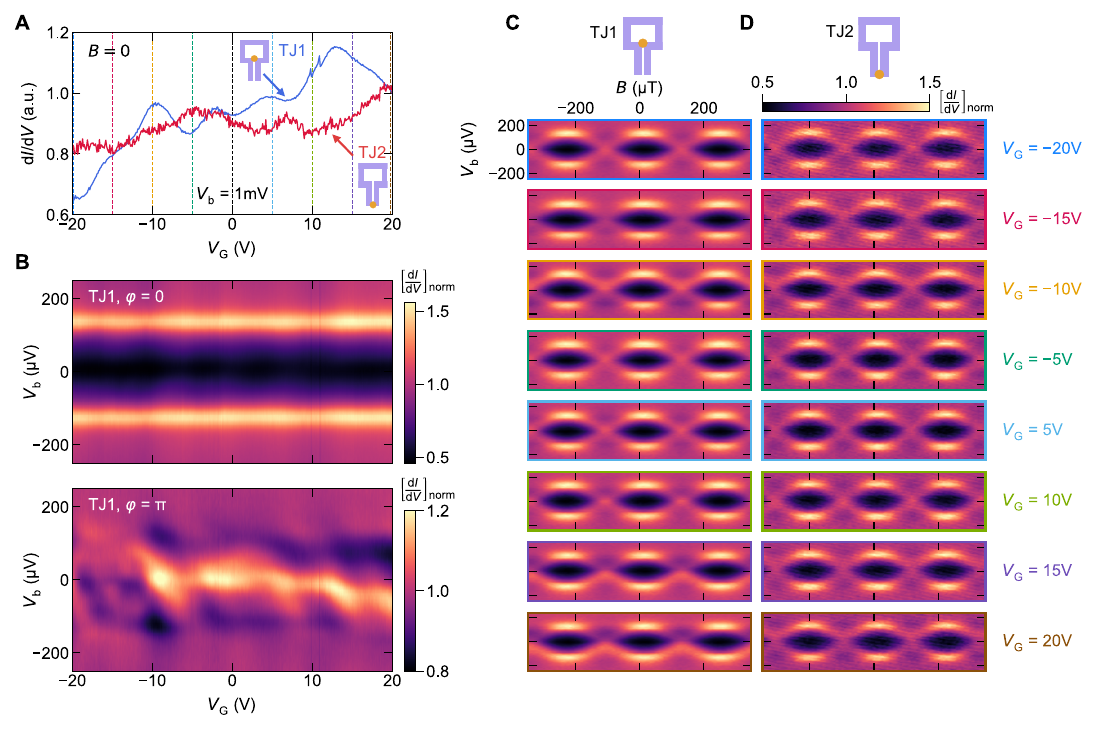}
\caption{\linespread{1.05}\selectfont{}
{\bf Stability of the spectra against a change in the chemical potential.} 
\textbf{A,} Differential conductance at a high bias voltage of 1 mV as a function of globally-applied back-gate voltage $V_{\rm G}$. For both tunnel contacts TJ1 and TJ2, the overall conductance is reduced for more negative gate voltages, indicating an $n$-type regime. The vertical dashed lines denote the gate voltages for which the continuous evolution of the spectrum is shown in panels {\bf C} and {\bf D}. \textbf{B,} Normalized d$I$/d$V$, denoted $\left[\frac{\mathrm{d}I}{\mathrm{d}V}\right]_{\rm norm}$, for TJ1 mapped in the $V_{\rm b}$ vs $V_{\rm G}$ plane for $\varphi = 0$ (upper panel) and $\varphi =\pi$ (lower panel). The normalization was done at each $V_{\rm G}$ by using the d$I$/d$V$ value at $V_{\rm b} = 0.3$ mV as reference.
\textbf{C,D,} Evolution of the $B$-dependence of the spectrum for a wide range of $V_{\rm G}$ from $-20$ to $+20$ V  for TJ1 ({\bf C}) and TJ2 ({\bf D}). The $V_{\rm G}$ values selected for the presentation are equally spaced.}
\label{fig:VG}
\end{figure}

In the next step, we check the robustness of the observed features with respect to the chemical potential, as tuned by a global back gate. Figure~\ref{fig:VG}A shows the d$I$/d$V$ data in $B$ = 0 T at a high bias voltage ($V_{\rm b}$ = 1\,mV) as a function of the back-gate voltage $V_{\rm G}$ for both TJ1 and TJ2. The overall tendency is that the conductance decreases for more negative gate-voltages, indicating that the TSS is in the $n$-type regime. Superimposed to this general trend, we observe broad peaks (typical peak width of a few V), which are commonly found in dielectric-based tunneling devices and conventionally explained by resonant states in the dielectric \cite{Bretheau2017}. We can use their dispersion to estimate the relation between $V_{\rm G}$ and $\mu$ (see \ref{SI:HighVB}) and find $\Delta \mu / \Delta V_{\rm G} \approx 1-2\,\mathrm{meV/V}$ in the given range of chemical potentials, consistent with results reported in our earlier work \cite{Ghatak2018}. This implies that the chemical potential $\mu$ changes by at least $40\,\mathrm{meV}$ in the gating range studied here.

{Figure~\ref{fig:VG}B presents the $V_{\rm G}$-dependencies of the conductance spectra obtained at TJ1, normalized at each $V_{\rm G}$ by using the d$I$/d$V$ value at $V_{\rm b} = 0.3$ mV as reference. Here, we show the evolutions of the maximum excitation gap for $\varphi = 0$ (upper panel) and the spectrum observed for $\varphi = \pi$ (lower panel) in a wide range of $V_{\rm G}$ from $-20$ to $+20$ V. The maximum excitation gap is very stable, indicating that the resonance states due to impurities are not affecting the proximity-induced gap. 
The apparent gap closing at $\varphi = \pi$ is also very stable and is observed at all $V_{\rm G}$ values. Note that the occurrence of the broad ZBCP and the details of the LDOS near zero-bias appear to depend on $V_{\rm G}$, but as we noted above, these features are expected to be sensitive to the details of the experimental conditions; for example, the d$I$/d$V$ peak is slightly shifted to the negative $V_{\rm b}$ side at $V_{\rm G} \approx$ 12 -- 20 V, which is probably due to the strong resonance state located around 13 V (cf. Fig. \ref{fig:VG}A) crossing the gap energy. We speculate that this strongly breaks particle-hole symmetry, resulting in the shift of the peak from zero to finite energy. What is important in these data is the robust appearance of highly transparent ABSs, potentially indicating a gap closing, which is observed for the whole range of $V_{\rm G}$.}

In Figs. \ref{fig:VG}C and \ref{fig:VG}D, we show the $B$-dependencies of the normalized conductance spectra for equally-spaced $V_{\rm G}$ values between $-20$ and $+20$ V for both TJ1 and TJ2. One can immediately see that the basic magnetic-field-periodic change in d$I$/d$V$ is very stable at all $V_{\rm G}$ values. Upon close inspection, the breaking of exact electron-hole symmetrical conductance, as mentioned above, is found to be modified by the back-gate voltage. Similar behavior has been reported for comparable devices, e.g. based on graphene or semiconductors \cite{Bretheau2017,Fornieri2018}.
Overall, we conclude that the apperance of modes with high transmission is observed irrespective of changes of the chemical potential, possibly indicating the TSS origin. This is further supported by a device with a wider range of gate-tunability, showing the same phenomenology (see \ref{SI:VgX}).  

\section{Discussion}
Whether our observations manifest a gap closing and reopening of the proximitized TSS and therefore demonstrate a topological phase transition is the crucial question for the interpretation of our experimental data. From spectroscopy of individual ABSs (see below and \ref{SI:QDs}), we can estimate the energy resolution of our setup to be about 30 $\upmu$eV. According to Eq. (1), when $\Delta \approx$ 130 $\upmu$eV and the experimental energy resolution is 30 $\upmu$eV, a spectrum with a mini-gap would become indistinguishable from a gapless spectrum for $\tau \geq 0.95$ . This places a lower bound on the transmission value that is required to observe an apparent gap closing in our experimental setup.  
The fact that we consistently observe signatures of such highly transmissive modes, makes an interpretation in terms of topologically protected $\tau=1$ modes plausible, but not inevitable. In the following we also want to discuss topologically trivial scenarios that could result in an apparent gap closing at $\varphi = \pi$.
A quantum-dot state, symmetrically coupled to SC electrodes and tuned to resonance, leads to a zero-energy state \cite{Oriekhov2021}. This highly fine-tuned case was in fact observed by us in a device where the tunnel barrier was not in a favorable condition to probe the TSS. We found strongly dispersing bound states in a very narrow range of $V_{\rm G}$ associated with a spurious ZBCP occurring at a particular $V_{\rm G}$ value (see \ref{SI:QDs}). These features are consistent with the Andreev quantum-dot states \cite{Oriekhov2021} which might occur in the dielectric. The robust spectra discussed in the previous section are clearly of different origin. 
The spurious ZBCP in this case also tells us that one should not look for a ZBCP but should focus on the gap closing to elucidate the robustness of the phenomenon.

In the literature, the gap closing at $\varphi = \pi$ has been reported for diffusive long SNS junctions \cite{Zhou1998,leSueur2008}. In such a system, disorder localizes most channels, but it also leads to the appearance of a few high-transmission channels \cite{Beenakker1997}, which can be responsible for a gap closing. 

In the case of our junctions, the mean-free path $l_{\rm e} \approx$ 40 nm has been estimated for the related material of bulk-insulating Bi$_{1.5}$Sb$_{0.5}$Te$_{1.7}$Se$_{1.3}$ crystals \cite{Taskin2011}, which is a bit shorter but comparable to our junction length (50 nm), putting our device in a quasi-ballistic regime. (Remember that our device is in the short-junction limit where the Fu-Kane theory is applicable.) It is possible that some nearly-longitudinal Andreev modes with large $k_y$ traveling for a long distance present the gap closing phenomenology of the diffusive scenario, although we consider it unlikely that this should be as robust against gating as we observe, in particular, as the number of modes in our TI device should be significantly smaller compared to previous experiments on normal metals like Ag\cite{leSueur2008}.

On the other hand, even if a more refined experiment would find a finite gap in the spectra, this would not definitely rule out the existence of the topologically protected $\tau=1$ mode. As has recently been discussed \cite{Kim2025}, in realistic device geometries, the coupling of the modes at the boundary to the ones in the junctions can lift the energies from zero. Similarly, if the wave functions of the ABSs underneath the superconducting electrons do not decay rapidly enough, they could hybridize with TI states outside of the proximitized region. 
This effect could be overcome by choosing a different device geometry like the trijunction geometry\cite{Fu2008} or the nano-SQUID recently proposed by us\cite{Nikodem2025}.  

To make more definite statements about the $\tau$ values of individual states and to decide whether a gap closing takes place, employing high-frequency measurements to achieve a higher resolution \cite{Murani2019} would be useful. In this context it will also be important to obtain better control over the chemical potential in the TI, to control the number of modes in the junction and exclude trivial scenarios. Furthermore, employing a charge-sensing technique \cite{Ben-Shach2015} to detect the parity change in the junction before and after the gap closing would unambiguously prove the topological phase transition.

A definitive proof of Majorana zero-modes is the demonstration of their non-Abelian exchange statistics through braiding. In a TI Josephson junction, well-protected Majorana zero-modes can in principle be generated in Josephson vortex-cores as we discuss below. Once such protected Majorana zero-modes are generated, they can be physically moved along the junction in a controlled manner \cite{Hedge2020}; this degree of freedom allows us to conceive braiding operations in a three-slit Josephson junction \cite{Hedge2020} and the resulting change in the parity can be detected by charge sensing \cite{Ben-Shach2015}. Another possibility to confirm the non-Abelian statistics is to build a Josephson-vortex interferometer proposed by Grosfeld and Stern \cite{Grosfeld2011}, which detects the result of braiding using the Aharonov-Casher effect.

\begin{figure*}
\centering
\includegraphics[width=\textwidth]{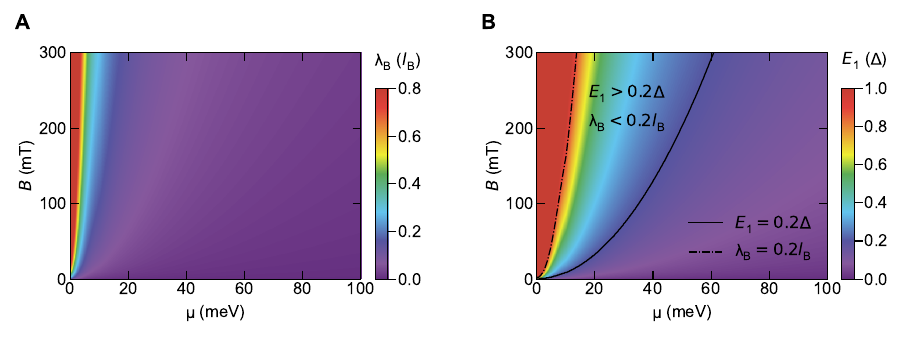}
{\caption{\linespread{1.05}\selectfont{}
{\bf Stability of the Majorana zero-modes in the putative case with Nb electrodes.} 
\textbf{A,B,} Calculated dependencies of the Majorana bound state width $\lambda_{\rm B}$ and the first excited state energy $E_1$ on the chemical potential $\mu$ and the perpendicular magnetic field $B$ for the Majorana zero-mode generated at the Josephson vortex-core in a putative device with Nb as the SC electrodes, which will sustain high magnetic fields. We used $L$ = 50 nm, $v_{\rm F}$ = 5.5 $\times$ 10$^5$ m/s, and $\Delta$ = 1 meV. The dashed and solid black lines in panel {\bf B} are the boundaries to guarantee a sufficient separation of Majorana zero-modes ($\lambda_{\rm B}/l_{\rm B} < 0.2$ obtained from panel {\bf A}) and a sufficiently large excitation gap ($E_1/\Delta > 0.2$), respectively. Majorana zero-modes are well-protected in the relatively broad region between the dashed and solid black lines. 
}
\label{fig:Nb}}
\end{figure*}

Let us estimate the necessary parameter range to obtain well-protected Majorana zero-modes in Josephson vortex-cores. There are two quantities that are crucial for this consideration: One is the spatial extent of the Majorana zero-mode $\lambda_{\rm B} = \sqrt{v_{\rm M} l_{\rm B}/\pi \Delta}$, where $l_{\rm B} = \Phi_0/(B L_{\rm eff})$ is the distance between Josephson vortices in the junction and $v_{\rm M}$ is the renormalized Majorana velocity. To calculate $v_{\rm M}$, we use the formula (see \ref{SI:vm})
\begin{equation}
v_{\rm M} = v_{\rm F}\left(\frac{\Delta^2}{\Delta^2 + \mu^2}\right) \frac{\cos\left(\frac{\mu L}{\hbar v_{\rm F}}\right) + \frac{\Delta}{\mu} \sin\left(\frac{\mu L}{\hbar v_{\rm F}}\right)}{1 + \frac{\Delta L}{\hbar v_{\rm F}}}
\end{equation}
valid in the short-junction regime $L \lesssim \hbar v_{\rm F}/\Delta$ ($L$ is the true length of the junction). For large $\mu\gg \Delta$, the formula goes over to the simple expression $v_{\rm M} \simeq v_{\rm F}(\Delta/\mu)^2$ given in Ref. \citenum{Potter2013}. The other important quantity is the energy of the first excited state, which is given as $E_1 \approx \sqrt{2\pi v_{\rm M}\Delta/l_{\rm B}}$ near $\varphi = (2n+1)\pi$  \cite{Potter2013}. 
Note that reducing $\mu$ enhances $v_{\rm M}$, and thus $E_1$ and $\lambda_{\rm B}$. However, $\lambda_{\rm B}$ must be shorter than $l_{\rm B}/2$ to avoid the overlap of neighboring Majorana zero-modes which bounds $\mu$ from below. To see this competition, we theoretically consider the case of using Nb for the SC electrodes, since its higher magnetic field tolerance allows us to study a wider parameter range as compared to Al. We calculate the dependencies of $\lambda_{\rm B}$ and $E_1$ on $\mu$ and $B$. The results, shown in {Figs.~\ref{fig:Nb}A and \ref{fig:Nb}B}, indicate that a moderately large $\mu$ is indeed helpful to realize a short $\lambda_{\rm B}$ and the magnetic field helps to obtain a usefully large $E_1$. In {Fig.~\ref{fig:Nb}B}, we draw the boundaries to guarantee a sufficient separation of Majorana zero-modes ($\lambda_{\rm B}/l_{\rm B} < 0.2$, dash-dotted line) and a sufficiently large excitation gap ($E_1/\Delta > 0.2$, solid line). In the region between the lines, see {Fig.~\ref{fig:Nb}B},  Majorana zero-modes are well-protected. In fact, these estimates suggest that it is not necessary to tune $\mu$ very close to the Dirac point as long as a magnetic field of the order of 100 mT is used to generate Josephson vortices --- an encouraging result for this platform. Possible complications, however, could arise from local variations of the chemical potential as well as Abrikosov vortices generated in the superconducting electrodes at these magnetic field scales, possibly resulting in unstable phase profiles and magnetic field distributions. Furthermore, it is prudent to note that the soft gap, which we observed in our devices, is a problem for braiding, and a hard gap must be realized by improving the interface transparency between the TI and a superconductor and potentially optimizing the device geometry to eliminate long trajectories\cite{Laeven2020}. 
  
In conclusion, we have observed a periodic modulation of the density of states in a phase-biased TI Josephson junction, which could indicate the periodic gap closing and reopening expected from the existence of perfectly-transmitted Andreev modes guaranteed by the spin-momentum locking of the TSS. This phenomenon is consistently observed in several devices as a function of an extended parameter range in the chemical potential and the perpendicular magnetic field, which support its TSS origin. The spin-non-degeneracy of the TSS would make this a topological phase transition accompanied by a parity change, which remains to be confirmed by future experiments. In the magnetic-field range where a couple of Josephson vortices nucleate in the junction, we observed the locality of the spectra implying a confinement of the zero-energy mode \cite{Potter2013}. However, within the magnetic-field range where Al remains superconducting, an abundance of low-lying excited states arises and this leads to spectra that are indistinguishable from those near zero field. We have discussed the necessary parameter range to realize well-protected Majorana zero-modes in the Josephson vortex core. Intriguingly, we found that maintaining a chemical potential $\mu$ not too close to the Dirac point is crucial for avoiding Majorana delocalization; a $\mu$ of approximately 25~meV facilitates the realization of well-protected Majorana zero-modes in a magnetic field of 100~mT accessible with a Nb electrode. Considering the high controllability of the Josephson vortex, this result is promising for braiding experiments. Overall, our experiment demonstrates the highly promising nature of the TI Josephson junctions for the Majorana research thanks to the naturally-existing spin-momentum locking of the TSS.

\section{Materials and Methods}

\textbf{Materials and device fabrication:}
Bulk single crystals of BiSbTeSe$_2$, with nominal stoichiometry of Bi:Sb:Te:Se = 1:1:1:2, were synthesized by the modified Bridgman method using high-purity (99.9999\%) Bi, Sb, Te, and Se as starting materials by following the recipe described in Ref.~ \citenum{Ren2011}.  We mechanically exfoliated thin (10-20 nm) BiSbTeSe$_2$ flakes under atmospheric conditions from bulk single crystals on degenerately-doped Si wafers covered by 280 nm SiO$_2$. The latter acts as dielectric for the back gate. Before the exfoliation, the substrates were cleaned with acetone, isopropanol and a low-power ozone treatment. BiSbTeSe$_2$ flakes suitable for device fabrication were identified under an optical microscope. The junctions were defined by electron beam lithography. PMMA A4 resist was exposed using a Raith PIONEER Two system. After development, residual resist in the contact areas was removed by a gentle oxygen plasma treatment. Subsequent cleaning in a hydrochloric acid solution shortly before metallization ensured the removal of any oxide layer. Then, 3~nm + 45~nm Ti/Al contacts were thermally evaporated after a gentle in-situ Ar cleaning to remove light atmospheric adsorbates and the device was exposed to air to form a native Al$_2$O$_3$ layer. Due to the bulk-insulating nature of our BiSbTeSe$_2$, the superconducting proximity effect only affects the surface in direct contact with the superconductor. An additional atomic layer deposition of a thin ($\leq\SI{1}{nm}$) Al$_2$O$_3$ layer onto the whole device was performed at 80 $^{\circ}$C using Ultratec Savannah S200 to form a tunnel barrier. For the fabrication of the tunnel electrodes, another electron beam lithography was performed to define the pattern. A 5-nm-thick Pt layer was sputter-deposited to act as wetting layer, followed by sputtering an additional 45-nm-thick Au layer. The contact resistances achieved by this procedure are typically larger than $100 \mathrm{kohms}$, ensuring electron tunneling to be the prevalent form of transport. 
While we can reliably produce Josephson junctions, the success rate of the tunnel junctions on top of them is less than 50\%, making it challenging to fabricate devices with two working tunnel contacts. The main reason for this is the very small required thickness of our dielectric, making it very sensitive to variations in both directions.
The precise device geometry was determined by scanning electron microscopy after the measurements were finished. 

\vspace{2mm}
\textbf{Measurements:}
Transport measurements were performed at base temperature ($\sim$50 mK) of our dry dilution refrigerator (Oxford Instruments TRITON 300). Electrical lines incorporate RC and copper-powder filters to reduce noise. Using a two-probe configuration, the differential conductance $\mathrm{d}I/\mathrm{d}V$  was measured between each individual tunnel electrode and the SC electrode with a standard low-frequency lock-in technique using a Basel Instruments current preamplifier. A small DC-offset introduced by the preamplifier was numerically corrected for after the measurements. An AC voltage of 5--10 $\upmu$V was superimposed to a DC bias voltage. For investigations of the magnetic field dependence, we used our integrated 6/1/1-T superconducting vector magnet.
\\

\vspace{2mm}
\textbf{Simulations:}
To qualitatively determine the differential conductance  of the TI Josephson junction, a transport simulation for a three-dimensional tight-binding experiment was performed using the Python library Kwant \cite{Groth2014}.
The simulation of a three-dimensional bulk TI is necessary to circumvent the Nielsen-Ninomiya (fermion-doubling) theorem~\cite{Nielsen1983}. The three-dimensional simulation allows us to realize a single Dirac-cone on each face of the TI, with the fermion-doubled partner on the opposite surface, where it can be gapped away by applying magnetic fields or opening a superconducting gap~\cite{Ziesen2023}.
As a model for the three-dimensional strong TI, the Bernevig-Hughes-Zhang (BHZ) model was chosen and implemented on a cubic lattice.
Superconductivity was included in the Bogoliubov--de-Gennes formalism. 
The system has been extended infinitely along the junction and probed by a tunnel lead attached perpendicularly to the junction.
In the basis $\boldsymbol{\Psi} = ((\Psi_\uparrow^1, \Psi_\downarrow^1),(\Psi_\uparrow^2, \Psi_\downarrow^2))^T$, where $i=1,2$ is denoting the orbital and $\uparrow, \downarrow$ the spin, the full BHZ Hamiltonian is given by
\begin{equation}\label{eq:BHZ_OG}
    H_\textrm{BHZ} = (\epsilon_0(\boldsymbol{k}) - \mu) + M(\boldsymbol{k}) \, \hat{\pi}_z + A_1 \, k_z \, \hat{\pi}_x \hat{\sigma}_z + A_2 (k_x \hat{\pi}_x \hat{\sigma}_x + k_y \, \hat{\pi}_x \hat{\sigma}_y)
\end{equation}
with
\begin{align}
    \epsilon_0(\boldsymbol{k}) &= C + D_1 k_z^2 + D_2 (k_x^2+k_y^2) \quad \textrm{and} \\
    M(\boldsymbol{k}) &= M - B_1 k_z^2 - B_2 (k_x^2+k_y^2) \, .
\end{align}
The Pauli operators $\boldsymbol{\hat{\pi}}$ act on the orbital degree of freedom and $\boldsymbol{\hat{\sigma}}$ on the spin degree of freedom.
In our notation, tensor products of different Pauli operators are always implied, and identity operators acting are not written explicitly.
For our purposes we use the simplest version of the model in Eq. (\ref{eq:BHZ_OG}) that exhibits a topological phase.
We neglect the spin and orbital independent part of the spectrum $\epsilon_0(\boldsymbol{k})$ by setting $C=D_j=0$ and choose 
\begin{equation}
    A_1 = A_2 = v \quad \textrm{and} \quad B_1 = B_2 = \frac{v}{2|M|} \, .
\end{equation}
With this choice of parameters, the Hamiltonian simplifies to
\begin{equation}\label{eq:BHZ}
    H_{\textrm{BHZ}} = -\mu + (M-\frac{v}{2|M|} \, \boldsymbol{k}^2) \hat{\pi}_z + v \hat{\pi}_x \, \boldsymbol{k} \cdot \boldsymbol{\hat{\sigma}} \, .
\end{equation}
The model exhibits a topological gap $2|M|$ and a Dirac surface spectrum with the Dirac velocity $v$, which is set to $1$ in the simulation.
Furthermore, the parameter $M$ was set to $1$.
The reduced model is implemented as a tight-binding model on a cubic lattice in Kwant.
The full simulation of the bulk of the TI is necessary to obtain correct surface states.
For the inclusion of superconductivity in the system, the time reversed states of the BHZ model were implemented as additional orbitals to explicitly simulate a Bogoliubov--de-Gennes (BdG) Hamiltonian.
On the surface of the TI, a finite superconducting gap $\Delta=0.1e^{i\varphi}$ is opened by coupling the two sectors of the BdG-Hamiltonian.
Two superconducting regions are separated by a bare TI surface of fixed width on the top of the sample, thus modeling a Josephson junction of the surfaces by varying $\varphi$ across the junction.
Only on that strip of bare TI surface the gapless surface state are found.
On all other surfaces, the surface states are gapped away by the finite superconductivity.
The phase variation created by the magnetic field in the experiment is implemented directly.
On the bottom surface, the phase jump is split up into two steps to prevent hybridization between the surface state in the TI JJ and unphysical Andreev bound state on the lower surface of the sample.
The cross section geometry is sketched in Fig.~\ref{fig:SI5}. 
The dimensions of the simulation in units of the lattice constant are $L=9$, $L_\text{SC}=8$, $L'_\text{SC}=1$, $L_{\rm T}=9$, $D=12$ and $D_\text{SC}=1$.
The system is extended infinitely along the junction and probed by a tunnel lead attached vertically to the bare TI region at one point.
The tunnel lead is simulated as a normal conducting cubic lattice and the tunnel contact is tuned by varying the hopping from the lead into the system.
The differential conductivity is calculated as a function of the energy $E$ compared to the superconducting gap $\Delta$ and the phase difference across the Josephson junction $\phi_B$.

\clearpage 

\bibliography{Bibliography} 

@article{Brede2024,
   author = {Brede, Jens and Bagchi, Mahasweta and Greichgauer, Adrian and Uday, Anjana and Bliesener, Andrea and Lippertz, Gertjan and Yazdanpanah, Roozbeh and Taskin, Alexey and Ando, Yoichi},
   title = {Characterizing the chemical potential disorder in the topological insulator ${({\mathrm{Bi}}_{1\ensuremath{-}x}{\mathrm{Sb}}_{x})}_{2}{\mathrm{Te}}_{3}$ thin films},
   journal = {Phys. Rev. Mater.},
   volume = {8},
   number = {10},
   pages = {104202},
   DOI = {10.1103/PhysRevMaterials.8.104202},
   year = {2024},
   type = {Journal Article}
}

@article{Laeven2020,
  title = {Enhanced Proximity Effect in Zigzag-Shaped Majorana Josephson Junctions},
  author = {Laeven, Tom and Nijholt, Bas and Wimmer, Michael and Akhmerov, Anton R.},
  journal = {Phys. Rev. Lett.},
  volume = {125},
  issue = {8},
  pages = {086802},
  numpages = {5},
  year = {2020},
  month = {Aug},
  publisher = {American Physical Society},
  doi = {10.1103/PhysRevLett.125.086802},
  opturl = {https://link.aps.org/doi/10.1103/PhysRevLett.125.086802}
}

@article{Taskin2011,
   author = {Taskin, A. A. and Ren, Zhi and Sasaki, Satoshi and Segawa, Kouji and Ando, Yoichi},
   title = {Observation of {D}irac Holes and Electrons in a Topological Insulator},
   journal = {Phys. Rev. Lett.},
   volume = {107},
   number = {1},
   DOI = {10.1103/PhysRevLett.107.016801},
   year = {2011},
   type = {Journal Article}
}

@article{Akhmerov2011,
	author={A. R. Akhmerov and J. P. Dahlhaus and F. Hassler and M. Wimmer and C. W. J. Beenakker},
	journal={Phys. Rev. Lett.},
	volume={106},
	pages={057001},
	doi={10.1103/PhysRevLett.106.057001},
	year={2011},
	title={Quantized conductance at the Majorana phase transition in a disordered superconducting wire}
}

@article{Rosdahl2018,
	title = {Andreev rectifier: {A} nonlocal conductance signature of topological phase transitions},
	volume = {97},
	issn = {24699969},
	doi = {10.1103/PhysRevB.97.045421},
	number = {4},
	journal = {Phys. Rev. B},
	author = {Rosdahl, T. \"O. and Vuik, A. and Kjaergaard, M. and Akhmerov, A. R.},
	year = {2018},
   pages = {045421},
}

@article{Xu2015,
   author = {Xu, Jin-Peng and Wang, Mei-Xiao and Liu, Zhi Long and Ge, Jian-Feng and Yang, Xiaojun and Liu, Canhua and Xu, Zhu An and Guan, Dandan and Gao, Chun Lei and Qian, Dong and Liu, Ying and Wang, Qiang-Hua and Zhang, Fu-Chun and Xue, Qi-Kun and Jia, Jin-Feng},
   title = {Experimental Detection of a {M}ajorana Mode in the core of a Magnetic Vortex inside a Topological Insulator-Superconductor {${\mathrm{Bi}}_{2}{\mathrm{Te}}_{3}/{\mathrm{NbSe}}_{2}$} Heterostructure},
   journal = {Phys. Rev. Lett.},
   volume = {114},
   number = {1},
   pages = {017001},
   year = {2015},
   type = {Journal Article}
}

@article{Ben-Shach2015,
   author = {Ben-Shach, Gilad and Haim, Arbel and Appelbaum, Ian and Oreg, Yuval and Yacoby, Amir and Halperin, Bertrand I.},
   title = {Detecting {M}ajorana modes in one-dimensional wires by charge sensing},
   journal = {Phys. Rev. B},
   volume = {91},
   number = {4},
   pages = {045403},
   year = {2015},
   type = {Journal Article}
}

@article{Hedge2020,
   author = {Hegde, Suraj S. and Yue, Guang and Wang, Yuxuan and Huemiller, Erik and Van Harlingen, D. J. and Vishveshwara, Smitha},
   title = {A topological {J}osephson junction platform for creating, manipulating, and braiding {M}ajorana bound states},
   journal = {Ann. Phy.},
   volume = {423},
   pages = {168326},
   ISSN = {0003-4916},
   DOI = {https://doi.org/10.1016/j.aop.2020.168326},
   year = {2020},
   type = {Journal Article}
}

@article{Oriekhov2021,
   author = {Oriekhov, D. O. and Cheipesh, Y. and Beenakker, C. W. J.},
   title = {Voltage staircase in a current-biased quantum-dot {J}osephson junction},
   journal = {Phys. Rev. B},
   volume = {103},
   number = {9},
   pages = {094518},
   DOI = {10.1103/PhysRevB.103.094518},
   year = {2021},
   type = {Journal Article}
}

@article{Sato2017,
   author = {Sato, Masatoshi and Ando, Yoichi},
   title = {Topological superconductors: a review},
   journal = {Rep. Prog. Phys.},
   volume = {80},
   number = {7},
   pages = {076501},
   ISSN = {0034-4885},
   DOI = {10.1088/1361-6633/aa6ac7},
   year = {2017},
   type = {Journal Article}
}

@article{Deacon2017,
   author = {Deacon, R. S. and Wiedenmann, J. and Bocquillon, E. and Dominguez, F. and Klapwijk, T. M. and Leubner, P. and Brune, C. and Hankiewicz, E. M. and Tarucha, S. and Ishibashi, K. and Buhmann, H. and Molenkamp, L. W.},
   title = {Josephson Radiation from Gapless {A}ndreev Bound States in {HgTe}-Based Topological Junctions},
   journal = {Phys. Rev. X},
   volume = {7},
   number = {2},
   pages = {021011},
   year = {2017},
   type = {Journal Article}
}

@article{Nayak2008,
  title = {Non-Abelian anyons and topological quantum computation},
  author = {Nayak, Chetan and Simon, Steven H. and Stern, Ady and Freedman, Michael and Das Sarma, Sankar},
  journal = {Rev. Mod. Phys.},
  volume = {80},
  issue = {3},
  pages = {1083--1159},
  numpages = {0},
  year = {2008},
  month = {Sep},
  publisher = {American Physical Society},
  doi = {10.1103/RevModPhys.80.1083},
  opturl = {https://link.aps.org/doi/10.1103/RevModPhys.80.1083}
}

@article{Kurter2013,
  title={Evidence for an anomalous current-phase relation in topological insulator {J}osephson junctions},
  author={Cihan Kurter and A. D. K. Finck and Yew San Hor and Dale J. van Harlingen},
  journal={Nat. Commun.},
  year={2013},
  volume={6},
  opturl={https://api.semanticscholar.org/CorpusID:29208010}
}

@article{Ghatak2018,
	author = {Ghatak, Subhamoy and Breunig, Oliver and Yang, Fan and Wang, Zhiwei and Taskin, Alexey A. and Ando, Yoichi},
	doi = {10.1021/acs.nanolett.8b02029},
	issn = {1530-6984},
	journal = {Nano Lett.},
	number = {8},
	pages = {5124-5131},
	title = {Anomalous {F}raunhofer Patterns in Gated {J}osephson Junctions Based on the Bulk-Insulating Topological Insulator {BiSbTeSe$_2$}},
	type = {Journal Article},
	volume = {18},
	year = {2018},
	bdsk-opturl-1 = {https://doi.org/10.1021/acs.nanolett.8b02029}
	}

@article{Potter2013,
  title = {Anomalous supercurrent from {M}ajorana states in topological insulator Josephson junctions},
  author = {Potter, Andrew C. and Fu, Liang},
  journal = {Phys. Rev. B},
  volume = {88},
  issue = {12},
  pages = {121109},
  numpages = {5},
  year = {2013},
  month = {Sep},
  publisher = {American Physical Society},
  doi = {10.1103/PhysRevB.88.121109},
  opturl = {https://link.aps.org/doi/10.1103/PhysRevB.88.121109}
}

@article{Shatz2014,
    doi = {10.1371/journal.pone.0111643},
    author = {Shatz, Lisa F. AND Christensen, Craig W.},
    journal = {PLOS ONE},
    publisher = {Public Library of Science},
    title = {Numerical Inductance Calculations Based on First Principles},
    year = {2014},
    month = {11},
    volume = {9},
    opturl = {https://doi.org/10.1371/journal.pone.0111643},
    pages = {1-8},
    number = {11},

}

@Article{Melo2021,
	title={{Conductance asymmetries in mesoscopic superconducting devices due to finite bias}},
	author={André Melo and Chun-Xiao Liu and Piotr Rożek and Tómas Örn Rosdahl and Michael Wimmer},
	journal={SciPost Phys.},
	volume={10},
	pages={037},
	year={2021},
	publisher={SciPost},
	doi={10.21468/SciPostPhys.10.2.037},
	opturl={https://scipost.org/10.21468/SciPostPhys.10.2.037},
}

@article{Stenger2017,
  title = {Tunneling conductance in semiconductor-superconductor hybrid structures},
  author = {Stenger, John and Stanescu, Tudor D.},
  journal = {Phys. Rev. B},
  volume = {96},
  issue = {21},
  pages = {214516},
  numpages = {15},
  year = {2017},
  month = {Dec},
  publisher = {American Physical Society},
  doi = {10.1103/PhysRevB.96.214516},
  opturl = {https://link.aps.org/doi/10.1103/PhysRevB.96.214516}
}

@article{Fu2008,
  title = {Superconducting Proximity Effect and {M}ajorana Fermions at the Surface of a Topological Insulator},
  author = {Fu, Liang and Kane, C. L.},
  journal = {Phys. Rev. Lett.},
  volume = {100},
  issue = {9},
  pages = {096407},
  numpages = {4},
  year = {2008},
  month = {Mar},
  publisher = {American Physical Society},
  doi = {10.1103/PhysRevLett.100.096407},
  opturl = {https://link.aps.org/doi/10.1103/PhysRevLett.100.096407}
}

@article{Banerjee2023,
  title = {Control of {A}ndreev Bound States Using Superconducting Phase Texture},
  author = {Banerjee, Abhishek and Geier, Max and Rahman, M. Ahnaf and Sanchez, Daniel S. and Thomas, Candice and Wang, Tian and Manfra, Michael J. and Flensberg, Karsten and Marcus, Charles M.},
  journal = {Phys. Rev. Lett.},
  volume = {130},
  issue = {11},
  pages = {116203},
  numpages = {6},
  year = {2023},
  month = {Mar},
  publisher = {American Physical Society},
  doi = {10.1103/PhysRevLett.130.116203},
  opturl = {https://link.aps.org/doi/10.1103/PhysRevLett.130.116203}
}

@article{Banerjee2023b,
  title = {Local and Nonlocal Transport Spectroscopy in Planar {J}osephson Junctions},
  author = {Banerjee, A. and Lesser, O. and Rahman, M. A. and Thomas, C. and Wang, T. and Manfra, M. J. and Berg, E. and Oreg, Y. and Stern, Ady and Marcus, C. M.},
  journal = {Phys. Rev. Lett.},
  volume = {130},
  issue = {9},
  pages = {096202},
  numpages = {6},
  year = {2023},
  month = {Feb},
  publisher = {American Physical Society},
  doi = {10.1103/PhysRevLett.130.096202},
  opturl = {https://link.aps.org/doi/10.1103/PhysRevLett.130.096202}
}

@article{Banerjee2023c,
  title = {Signatures of a topological phase transition in a planar {J}osephson junction},
  author = {Banerjee, A. and Lesser, O. and Rahman, M. A. and Wang, H.-R. and Li, M.-R. and Kringh\o{}j, A. and Whiticar, A. M. and Drachmann, A. C. C. and Thomas, C. and Wang, T. and Manfra, M. J. and Berg, E. and Oreg, Y. and Stern, Ady and Marcus, C. M.},
  journal = {Phys. Rev. B},
  volume = {107},
  issue = {24},
  pages = {245304},
  numpages = {12},
  year = {2023},
  month = {Jun},
  publisher = {American Physical Society},
  doi = {10.1103/PhysRevB.107.245304},
  opturl = {https://link.aps.org/doi/10.1103/PhysRevB.107.245304}
}

@article{Martin2014,
  title = {Nonequilibrium theory of tunneling into a localized state in a superconductor},
  author = {Martin, Ivar and Mozyrsky, Dmitry},
  journal = {Phys. Rev. B},
  volume = {90},
  issue = {10},
  pages = {100508},
  numpages = {5},
  year = {2014},
  month = {Sep},
  publisher = {American Physical Society},
  doi = {10.1103/PhysRevB.90.100508},
  opturl = {https://link.aps.org/doi/10.1103/PhysRevB.90.100508}
}

@Article{Groth2014,
  author    = {Groth, Christoph W and Wimmer, Michael and Akhmerov, Anton R and Waintal, Xavier},
  title     = {{Kwant: a software package for quantum transport}},
  journal   = {New J. Phys.},
  year      = {2014},
  volume    = {16},
  number    = {6},
  pages     = {063065},
  month     = {jun},
  doi       = {10.1088/1367-2630/16/6/063065},
  publisher = {IOP Publishing},
}

@article{ Bretheau2017,
Author = {Bretheau, Landry and Wang, Joel I-Jan and Pisoni, Riccardo and Watanabe,
   Kenji and Taniguchi, Takashi and Jarillo-Herrero, Pablo},
Title = {Tunnelling spectroscopy of {A}ndreev states in graphene},
Journal = {Nat. Phys.},
Year = {2017},
Volume = {13},
Number = {8},
Pages = {756+},
Month = {AUG},
DOI = {10.1038/NPHYS4110}}

@article{Pillet2010, 
title={{A}ndreev bound states in supercurrent-carrying carbon nanotubes revealed},
 volume={6}, DOI={10.1038/nphys1811},
  number={12}, 
  journal={Nat. Phys.},
  author={Pillet, J-D. and Quay, C. H. and Morfin, P. and Bena, C. and Yeyati, A. Levy and Joyez, P.},
   year={2010}, 
   month={Nov}
   , pages={965-969}}

@article{Moehle2022,
author = {Moehle, Christian M. and Rout, Prasanna K. and Jainandunsing, Nayan A. and Kuiri, Dibyendu and Ke, Chung Ting and Xiao, Di and Thomas, Candice and Manfra, Michael J. and Nowak, Micha\l P. and Goswami, Srijit},
title = {Controlling {A}ndreev Bound States with the Magnetic Vector Potential},
journal = {Nano Lett.},
volume = {22},
number = {21},
pages = {8601-8607},
year = {2022},
doi = {10.1021/acs.nanolett.2c03130},}

@article{Ren2018,
  title={Topological superconductivity in a phase-controlled {J}osephson junction},
  author={Hechen Ren and Falko Pientka and Sean Hart and Andrew T. Pierce and Michael Kosowsky and Lukas Lunczer and Raimund Schlereth and Benedikt Scharf and Ewelina M. Hankiewicz and Laurens W. Molenkamp and Bertrand I. Halperin and Amir Yacoby},
  journal={Nature},
  year={2018},
  volume={569},
  pages={93 - 98},
}

@article{Fornieri2018,
  title={Evidence of topological superconductivity in planar {J}osephson junctions},
  author={A. Fornieri and A. M. Whiticar and F. Setiawan and El{\'i}as Portol{\'e}s Mar{\'i}n and Asbj{\o}rn C. C. Drachmann and Anna Keselman and Sergei Gronin and Candice Thomas and Tian Wang and Raymond L. Kallaher and Geoff C. Gardner and Erez Berg and Erez Berg and Michael J. Manfra and Ady Stern and Charles M. Marcus and Fabrizio Nichele},
  journal={Nature},
  year={2018},
  volume={569},
  pages={89 - 92},
}

@article{Kitaev2001,
doi = {10.1070/1063-7869/44/10S/S29},
opturl = {https://dx.doi.org/10.1070/1063-7869/44/10S/S29},
year = {2001},
month = {oct},
publisher = {},
volume = {44},
number = {10S},
pages = {131},
author = {Kitaev, A. Yu },
title = {Unpaired {M}ajorana fermions in quantum
wires},
journal = {Physics-Uspekhi},

}

@article{Aghaee2023,
  title = {{InAs-Al} hybrid devices passing the topological gap protocol},
  author = {Aghaee, Morteza and Akkala, Arun and Alam, Zulfi and Ali, Rizwan and Alcaraz Ramirez, Alejandro and Andrzejczuk, Mariusz and Antipov, Andrey E. and Aseev, Pavel and Astafev, Mikhail and Bauer, Bela and Becker, Jonathan and Boddapati, Srini and Boekhout, Frenk and Bommer, Jouri and Bosma, Tom and Bourdet, Leo and Boutin, Samuel and Caroff, Philippe and Casparis, Lucas and Cassidy, Maja and Chatoor, Sohail and Christensen, Anna Wulf and Clay, Noah and Cole, William S. and Corsetti, Fabiano and Cui, Ajuan and Dalampiras, Paschalis and Dokania, Anand and de Lange, Gijs and de Moor, Michiel and Estrada Salda\~na, Juan Carlos and Fallahi, Saeed and Fathabad, Zahra Heidarnia and Gamble, John and Gardner, Geoff and Govender, Deshan and Griggio, Flavio and Grigoryan, Ruben and Gronin, Sergei and Gukelberger, Jan and Hansen, Esben Bork and Heedt, Sebastian and Herranz Zamorano, Jes\'us and Ho, Samantha and Holgaard, Ulrik Laurens and Ingerslev, Henrik and Johansson, Linda and Jones, Jeffrey and Kallaher, Ray and Karimi, Farhad and Karzig, Torsten and King, Cameron and Kloster, Maren Elisabeth and Knapp, Christina and Kocon, Dariusz and Koski, Jonne and Kostamo, Pasi and Krogstrup, Peter and Kumar, Mahesh and Laeven, Tom and Larsen, Thorvald and Li, Kongyi and Lindemann, Tyler and Love, Julie and Lutchyn, Roman and Madsen, Morten Hannibal and Manfra, Michael and Markussen, Signe and Martinez, Esteban and McNeil, Robert and Memisevic, Elvedin and Morgan, Trevor and Mullally, Andrew and Nayak, Chetan and Nielsen, Jens and Nielsen, William Hvidtfelt Padk\ae{}r and Nijholt, Bas and Nurmohamed, Anne and O'Farrell, Eoin and Otani, Keita and Pauka, Sebastian and Petersson, Karl and Petit, Luca and Pikulin, Dmitry I. and Preiss, Frank and Quintero-Perez, Marina and Rajpalke, Mohana and Rasmussen, Katrine and Razmadze, Davydas and Reentila, Outi and Reilly, David and Rouse, Richard and Sadovskyy, Ivan and Sainiemi, Lauri and Schreppler, Sydney and Sidorkin, Vadim and Singh, Amrita and Singh, Shilpi and Sinha, Sarat and Sohr, Patrick and Stankevi\ifmmode \check{c}\else \v{c}\fi{}, Toma\ifmmode \check{s}\else \v{s}\fi{} and Stek, Lieuwe and Suominen, Henri and Suter, Judith and Svidenko, Vicky and Teicher, Sam and Temuerhan, Mine and Thiyagarajah, Nivetha and Tholapi, Raj and Thomas, Mason and Toomey, Emily and Upadhyay, Shivendra and Urban, Ivan and Vaitiek\ifmmode \dot{e}\else \.{e}\fi{}nas, Saulius and Van Hoogdalem, Kevin and Van Woerkom, David and Viazmitinov, Dmitrii V. and Vogel, Dominik and Waddy, Steven and Watson, John and Weston, Joseph and Winkler, Georg W. and Yang, Chung Kai and Yau, Sean and Yi, Daniel and Yucelen, Emrah and Webster, Alex and Zeisel, Roland and Zhao, Ruichen},
  collaboration = {Microsoft Quantum},
  journal = {Phys. Rev. B},
  volume = {107},
  issue = {24},
  pages = {245423},
  numpages = {54},
  year = {2023},
  month = {Jun},
  publisher = {American Physical Society},
  doi = {10.1103/PhysRevB.107.245423},
  opturl = {https://link.aps.org/doi/10.1103/PhysRevB.107.245423}
}

@misc{Pikulin2021,
      title={Protocol to identify a topological superconducting phase in a three-terminal device}, 
      author={Dmitry I. Pikulin and Bernard van Heck and Torsten Karzig and Esteban A. Martinez and Bas Nijholt and Tom Laeven and Georg W. Winkler and John D. Watson and Sebastian Heedt and Mine Temurhan and Vicky Svidenko and Roman M. Lutchyn and Mason Thomas and Gijs de Lange and Lucas Casparis and Chetan Nayak},
      year={2021},
      eprint={arXiv:2103.12217},
      archivePrefix={arXiv},
      primaryClass={cond-mat.mes-hall}
}

@article{Krogstrup2014,
  title={Epitaxy of semiconductor-superconductor nanowires.},
  author={Peter Krogstrup and N. L. B. Ziino and W. Chang and S. M. Albrecht and Morten Hannibal Madsen and Erik Johnson and Jesper Nyg{\aa}rd and Charles M. Marcus and Thomas Sand Jespersen},
  journal={Nat. Mater.},
  year={2014},
  volume={14 4},
  pages={
          400-6
        },
}

@article{Wiedenmann2016,
  title={4$\pi$-periodic {J}osephson supercurrent in {HgTe}-based topological {J}osephson junctions},
   author = {Wiedenmann, J. and Bocquillon, E. and Deacon, R. S. and Hartinger, S. and Herrmann, O. and Klapwijk, T. M. and Maier, L. and Ames, C. and Brune, C. and Gould, C. and Oiwa, A. and Ishibashi, K. and Tarucha, S. and Buhmann, H. and Molenkamp, L. W.},
   journal = {Nat. Commun.},
   volume = {7},
   pages = {10303},
   DOI = {10.1038/ncomms10303},
   year = {2016},
   type = {Journal Article}
}

@article{Ivanov2001,
  title = {Non-Abelian Statistics of Half-Quantum Vortices in $\mathit{p}$-Wave Superconductors},
  author = {Ivanov, D. A.},
  journal = {Phys. Rev. Lett.},
  volume = {86},
  issue = {2},
  pages = {268--271},
  numpages = {0},
  year = {2001},
  month = {Jan},
  publisher = {American Physical Society},
  doi = {10.1103/PhysRevLett.86.268},
  opturl = {https://link.aps.org/doi/10.1103/PhysRevLett.86.268}
}

@article{Read2000,
  title = {Paired states of fermions in two dimensions with breaking of parity and time-reversal symmetries and the fractional quantum Hall effect},
  author = {Read, N. and Green, Dmitry},
  journal = {Phys. Rev. B},
  volume = {61},
  issue = {15},
  pages = {10267--10297},
  numpages = {0},
  year = {2000},
  month = {Apr},
  publisher = {American Physical Society},
  doi = {10.1103/PhysRevB.61.10267},
  opturl = {https://link.aps.org/doi/10.1103/PhysRevB.61.10267}
}

@article{Ren2011,
  title = {Optimizing {Bi$_{2-x}$Sb$_x$Te$_{3-y}$Se$_y$} solid solutions to approach the intrinsic topological insulator regime},
  author = {Ren, Zhi and Taskin, A. A. and Sasaki, Satoshi and Segawa, Kouji and Ando, Yoichi},
  journal = {Phys. Rev. B},
  volume = {84},
  issue = {16},
  pages = {165311},
  numpages = {6},
  year = {2011},
  month = {Oct},
  publisher = {American Physical Society},
  doi = {10.1103/PhysRevB.84.165311},
  opturl = {https://link.aps.org/doi/10.1103/PhysRevB.84.165311}
}

@article{Arakane2012,
author = {Arakane, T and Sato, T and Souma, Serge and Kosaka, K and Nakayama, K and Komatsu, Masanori and Takahashi, T and Ren, Zhi and Segawa, Kouji and Ando, Yoichi},
year = {2012},
month = {01},
pages = {636},
title = {Tunable {D}irac cone in the topological insulator {Bi$_{2-x}$Sb$_x$Te$_{3-y}$Se$_y$}},
volume = {3},
journal = {Nat. Commun.},
doi = {10.1038/ncomms1639}
}

@article{Grosfeld2011,
author = {Eytan Grosfeld  and Ady Stern },
title = {Observing {M}ajorana bound states of Josephson vortices in topological superconductors},
journal = {Proc. Natl. Acad. Sci.},
volume = {108},
number = {29},
pages = {11810-11814},
year = {2011},
doi = {10.1073/pnas.1101469108},
opturl = {https://www.pnas.org/doi/abs/10.1073/pnas.1101469108},}

@article{Alicea2012,
  title={New directions in the pursuit of {M}ajorana fermions in solid state systems},
  author={Alicea, Jason},
  journal={Rep. Prog. Phys.},
  volume={75},
  number={7},
  pages={076501},
  year={2012},
  publisher={IOP Publishing}
}

@article{Dassarma2023,
  title={In search of {M}ajorana},
  author={Das Sarma, Sankar},
  journal={Nat. Phys.},
  volume={19},
  number={2},
  pages={165--170},
  year={2023},
  publisher={Nature Publishing Group UK London}
}

@article{Flensberg2021,
  title={Engineered platforms for topological superconductivity and {M}ajorana zero modes},
  author={Flensberg, Karsten and von Oppen, Felix and Stern, Ady},
  journal={Nat. Rev. Mater.},
  volume={6},
  number={10},
  pages={944--958},
  year={2021},
  publisher={Nature Publishing Group UK London}
}

@article{Hess2023,
  title = {Trivial {A}ndreev Band Mimicking Topological Bulk Gap Reopening in the Nonlocal Conductance of Long Rashba Nanowires},
  author = {Hess, Richard and Legg, Henry F. and Loss, Daniel and Klinovaja, Jelena},
  journal = {Phys. Rev. Lett.},
  volume = {130},
  issue = {20},
  pages = {207001},
  numpages = {6},
  year = {2023},
  month = {May},
  publisher = {American Physical Society},
  doi = {10.1103/PhysRevLett.130.207001},
  opturl = {https://link.aps.org/doi/10.1103/PhysRevLett.130.207001}
}

@article{Yue2024,
  title = {Signatures of {M}ajorana bound states in the diffraction patterns of extended superconductor--topological insulator--superconductor Josephson junctions},
  author = {Yue, Guang and Zhang, Can and Huemiller, Erik D. and Montone, Jessica H. and Arias, Gilbert R. and Wild, Drew G. and Zhang, Jered Y. and Hamilton, David R. and Yuan, Xiaoyu and Yao, Xiong and Jain, Deepti and Moon, Jisoo and Salehi, Maryam and Koirala, Nikesh and Oh, Seongshik and Van Harlingen, Dale J.},
  journal = {Phys. Rev. B},
  volume = {109},
  issue = {9},
  pages = {094511},
  numpages = {15},
  year = {2024},
  month = {Mar},
  publisher = {American Physical Society},
  doi = {10.1103/PhysRevB.109.094511},
  opturl = {https://link.aps.org/doi/10.1103/PhysRevB.109.094511}
}

@article{Williams2012,
  title = {Unconventional Josephson Effect in Hybrid Superconductor-Topological Insulator Devices},
  author = {Williams, J. R. and Bestwick, A. J. and Gallagher, P. and Hong, Seung Sae and Cui, Y. and Bleich, Andrew S. and Analytis, J. G. and Fisher, I. R. and Goldhaber-Gordon, D.},
  journal = {Phys. Rev. Lett.},
  volume = {109},
  issue = {5},
  pages = {056803},
  numpages = {5},
  year = {2012},
  month = {Jul},
  publisher = {American Physical Society},
  doi = {10.1103/PhysRevLett.109.056803},
  opturl = {https://link.aps.org/doi/10.1103/PhysRevLett.109.056803}
}

@article{Schuffelgen2019,
  title     = "Selective area growth and stencil lithography for in situ
               fabricated quantum devices",
  author    = "Sch{\"u}ffelgen, Peter and Rosenbach, Daniel and Li, Chuan and
               Schmitt, Tobias W and Schleenvoigt, Michael and Jalil, Abdur R
               and Schmitt, Sarah and K{\"o}lzer, Jonas and Wang, Meng and
               Bennemann, Benjamin and Parlak, Umut and Kibkalo, Lidia and
               Trellenkamp, Stefan and Grap, Thomas and Meertens, Doris and
               Luysberg, Martina and Mussler, Gregor and Berenschot, Erwin and
               Tas, Niels and Golubov, Alexander A and Brinkman, Alexander and
               Sch{\"a}pers, Thomas and Gr{\"u}tzmacher, Detlev",
  journal   = "Nat. Nanotechnol.",
  publisher = "Springer Science and Business Media LLC",
  volume    =  14,
  number    =  9,
  pages     = "825--831",
  month     =  sep,
  year      =  2019,
  language  = "en"
}

@article{Valentini2021,
  title={Nontopological zero-bias peaks in full-shell nanowires induced by flux-tunable {A}ndreev states},
  author={Valentini, Marco and Pe{\~n}aranda, Fernando and Hofmann, Andrea and Brauns, Matthias and Hauschild, Robert and Krogstrup, Peter and San-Jose, Pablo and Prada, Elsa and Aguado, Ram{\'o}n and Katsaros, Georgios},
  journal={Science},
  volume={373},
  number={6550},
  pages={82--88},
  year={2021},
  publisher={American Association for the Advancement of Science}
}

@article{Beenakker1991,
  title = {Universal limit of critical-current fluctuations in mesoscopic {J}osephson junctions},
  author = {Beenakker, C. W. J.},
  journal = {Phys. Rev. Lett.},
  volume = {67},
  issue = {27},
  pages = {3836},
  numpages = {0},
  year = {1991},
  month = {Dec},
  publisher = {American Physical Society},
  doi = {10.1103/PhysRevLett.67.3836},
}

@article{Dynes1978,
  title = {Direct Measurement of Quasiparticle-Lifetime Broadening in a Strong-Coupled Superconductor},
  author = {Dynes, R. C. and Narayanamurti, V. and Garno, J. P.},
  journal = {Phys. Rev. Lett.},
  volume = {41},
  issue = {21},
  pages = {1509--1512},
  numpages = {0},
  year = {1978},
  month = {Nov},
  publisher = {American Physical Society},
  doi = {10.1103/PhysRevLett.41.1509},
  opturl = {https://link.aps.org/doi/10.1103/PhysRevLett.41.1509}
}

@article{Lopez2023,
doi = {10.1088/1361-6668/adf360},
year = {2025},
month = {sep},
publisher = {IOP Publishing},
volume = {38},
number = {9},
pages = {095004},
author = {López-Núñez, David and Torras-Coloma, Alba and Portell-Montserrat, Queralt and Bertoldo, Elia and Cozzolino, Luca and Ummarino, Giovanni Alberto and Zaccone, Alessio and Rius, Gemma and Martínez, M and Forn-Díaz, P},
title = {Superconducting penetration depth of aluminum thin films},
journal = {Superconductor Science and Technology}}

@article{Zhang2010,
author = {Wei Zhang and Rui Yu and Hai-Jun Zhang and Xi Dai and Zhong Fang},
title = {First-principles studies of the three-dimensional strong topological insulators {Bi$_2$Te$_3$}, {Bi$_2$Se$_3$} and {Sb$_2$Te$_3$}},
journal = {New J. Phys.},
doi = {10.1088/1367-2630/12/6/065013},
year = {2010},
month = {jun},
publisher = {},
volume = {12},
number = {6},
pages = {065013},
}

@misc{AntonThesis,
  title        = {Master's thesis, {RWTH Aachen University}},
  author       = {Anton Montag},
  year         = 2023,
  note         = {Available at \url{https://doi.org/10.5281/zenodo.11615512}},
}

@article{leSueur2008,
  title = {Phase Controlled Superconducting Proximity Effect Probed by Tunneling Spectroscopy},
  author = {le Sueur, H. and Joyez, P. and Pothier, H. and Urbina, C. and Esteve, D.},
  journal = {Phys. Rev. Lett.},
  volume = {100},
  issue = {19},
  pages = {197002},
  numpages = {4},
  year = {2008},
  month = {May},
  publisher = {American Physical Society},
  doi = {10.1103/PhysRevLett.100.197002},
}

@article{Nichele2020,
  title = {Relating Andreev Bound States and Supercurrents in Hybrid Josephson Junctions},
  author = {Nichele, F. and Portol\'es, E. and Fornieri, A. and Whiticar, A. M. and Drachmann, A. C. C. and Gronin, S. and Wang, T. and Gardner, G. C. and Thomas, C. and Hatke, A. T. and Manfra, M. J. and Marcus, C. M.},
  journal = {Phys. Rev. Lett.},
  volume = {124},
  issue = {22},
  pages = {226801},
  numpages = {6},
  year = {2020},
  month = {Jun},
  publisher = {American Physical Society},
  doi = {10.1103/PhysRevLett.124.226801},
}

@article{Murani2019,
  title = {Microwave Signature of Topological {A}ndreev level Crossings in a Bismuth-based {J}osephson Junction},
  author = {Murani, A. and Dassonneville, B. and Kasumov, A. and Basset, J. and Ferrier, M. and Deblock, R. and Gu\'eron, S. and Bouchiat, H.},
  journal = {Phys. Rev. Lett.},
  volume = {122},
  issue = {7},
  pages = {076802},
  numpages = {6},
  year = {2019},
  month = {Feb},
  publisher = {American Physical Society},
  doi = {10.1103/PhysRevLett.122.076802},
}

@article{Beenakker1997,
  title={Random-matrix theory of quantum transport},
  author={Beenakker, C. W. J.},
  journal={Rev. Mod. Phys.},
  volume={69},
  number={3},
  pages={731},
  year={1997},
  publisher={APS}
}

@article{Zhou1998,
  title={Density of states in superconductor-normal metal-superconductor junctions},
  author={Zhou, F and Charlat, Ph and Spivak, B and Pannetier, B},
  journal={J. Low Temp. Phys.},
  volume={110},
  pages={841--850},
  year={1998},
  publisher={Springer}
}

@article{Tkachov2013,
  title = {Helical Andreev bound states and superconducting Klein tunneling in topological insulator Josephson junctions},
  author = {Tkachov, G. and Hankiewicz, E. M.},
  journal = {Phys. Rev. B},
  volume = {88},
  issue = {7},
  pages = {075401},
  numpages = {8},
  year = {2013},
  month = {Aug},
  publisher = {American Physical Society},
  doi = {10.1103/PhysRevB.88.075401},
  opturl = {https://link.aps.org/doi/10.1103/PhysRevB.88.075401}
}

@article{Ziesen2023,
  title = {Geometry-independent tight-binding method for massless Dirac fermions in two dimensions},
  author = {Ziesen, Alexander and Fulga, Ion Cosma and Hassler, Fabian},
  journal = {Phys. Rev. B},
  volume = {107},
  issue = {19},
  pages = {195409},
  numpages = {9},
  year = {2023},
  month = {May},
  publisher = {American Physical Society},
  doi = {10.1103/PhysRevB.107.195409},
  opturl = {https://link.aps.org/doi/10.1103/PhysRevB.107.195409}
}

@article{Nielsen1983,
  title={The Adler-Bell-Jackiw anomaly and Weyl fermions in a crystal},
  author={Nielsen, Holger Bech and Ninomiya, Masao},
  journal={Physics Letters B},
  volume={130},
  number={6},
  pages={389--396},
  year={1983},
  publisher={Elsevier},
  doi={https://doi.org/10.1016/0370-2693(83)91529-0}
}

@article{Kim2025,
  title = {Boundary-induced Majorana coupling in a planar topological Josephson junction},
  author = {Kim, Hyeongseop and Choi, Sang-Jun and Sim, H.-S. and Park, Sunghun},
  journal = {Phys. Rev. B},
  volume = {111},
  issue = {4},
  pages = {045414},
  numpages = {10},
  year = {2025},
  month = {Jan},
  publisher = {American Physical Society},
  doi = {10.1103/PhysRevB.111.045414},
 
}

@misc{Nikodem2025,
      title={Topological Insulator nano-{SQUID}: Flux-tunable platform for topological superconductivity}, 
      author={Ella Nikodem and Jakob Schluck and Henry F. Legg and Max Geier and Michal Papaj and Mahasweta Bagchi and Liang Fu and Yoichi Ando},
      year={2025},
      eprint={2412.07993},
      archivePrefix={arXiv},
      primaryClass={cond-mat.mes-hall},
      url={https://arxiv.org/abs/2412.07993}, 
}

@article{Pilgram2000,
  title = {Excitation spectrum of mesoscopic proximity structures},
  author = {Pilgram, S. and Belzig, W. and Bruder, C.},
  journal = {Phys. Rev. B},
  volume = {62},
  issue = {18},
  pages = {12462--12467},
  numpages = {0},
  year = {2000},
  month = {Nov},
  publisher = {American Physical Society},
  doi = {10.1103/PhysRevB.62.12462},

}
\bibliographystyle{sciencemag}

\section*{Acknowledgments:}
We thank Liang Fu, Max Geier, Micha{\l} Papaj, and Henry Legg  for fruitful discussions.
\paragraph*{Funding:}
YA has received funding from the European Research Council (ERC) under the European Union’s Horizon 2020 research and innovation program (Grant Agreement No. 741121) and from the Deutsche Forschungsgemeinschaft (DFG, German Research Foundation) under CRC 1238 - 277146847 (Subprojects A04 and B01). YA and FH have received funding from DFG under Germany's Excellence Strategy - Cluster of Excellence Matter and Light for Quantum Computing (ML4Q) EXC 2004/1 - 390534769.
\paragraph*{Author contributions:} 
YA conceived the project. MB grew the TI crystals. EN and JS fabricated the devices, performed the experiments and analyzed the data under the guidance of YA. AM and AZ performed numerical calculations under the guidance of FH. JS, FH, and YA wrote the manuscript with input from all authors.
\paragraph*{Competing Interests:} 
The authors declare they have no competing interests.
\paragraph*{Data, Code, and Materials Availability:}
All data and code needed to evaluate and reproduce the results in the paper are present in the paper and/or the Supplementary Materials.. To replicate the topological insulator crystals used in this study, follow the method in Ref. \cite{Ren2011} and the procedures outlined in the "Materials and device fabrication" section of the Materials and Methods. The data and codes that support the findings of this study are available at the online depository zenodo with the identifier {https://doi.org/10.5281/zenodo.11615512}. 
\subsection*{Supplementary materials}
Supplementary Text\\
Figs. S1 to S13\\

\newpage


\renewcommand{\thefigure}{S\arabic{figure}}
\renewcommand{\thetable}{S\arabic{table}}
\renewcommand{\theequation}{S\arabic{equation}}
\renewcommand{\thepage}{S\arabic{page}}
\setcounter{figure}{0}
\setcounter{table}{0}
\setcounter{equation}{0}
\setcounter{page}{1} 


\begin{center}
\section*{Supplementary Materials for\\ \scititle}

  Jakob~Schluck$^{\dagger\ast}$,
	Ella~Nikodem$^{\dagger}$,
	Anton~Montag,
	Alexander~Ziesen,\and
    Mahasweta~Bagchi,
    Fabian~Hassler,
    Yoichi~Ando$^{\ast}$\and\\
	\small$^\ast$Corresponding authors. Email: schluck@ph2.uni-koeln.de, ando@ph2.uni-koeln.de\and\\
	\small$^\dagger$These authors contributed equally to this work.
\end{center}

\subsubsection*{This PDF file includes:}

Supplementary Text\\
Figures S1 to S13\\

\newpage

\renewcommand{\thesection}{S\arabic{section}}
\renewcommand{\thesubsection}{S\arabic{section}.\arabic{subsection}}
\renewcommand{\thesubsubsection}{S\arabic{section}.\arabic{subsection}.\arabic{subsubsection}}

\setcounter{section}{0}

\section*{Supplementary Text}

\section{Critical temperature of the superconducting film}
\label{SI:TC}
\begin{figure}
\centering
\includegraphics{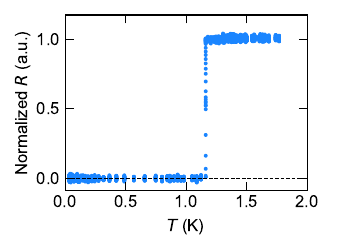}
\caption{\linespread{1.05}\selectfont{}
Temperature dependence of the resistance of the Al film measured in a four-probe geometry. The critical temperature is found to be $\approx 1.1$K.}
\label{fig:SIcooldown}
\end{figure}

In Fig. \ref{fig:SIcooldown}, we show a four-point resistance measurement of a pure aluminum film, which was deposited under the same conditions as the one used for the device fabrication. We see a sharp drop in resistance at $T_{\rm c}\approx1.15K$, indicating the superconducting phase transition. Via the standard BCS formula $\Delta=1.76k_{\rm B}T_{\rm c}$, this gives $\Delta\approx\SI{170}{\mu eV}$ for the aluminum film.

\section{Tunneling data of pure aluminum film}
\label{SI:BCS}
\begin{figure}
\centering
\includegraphics{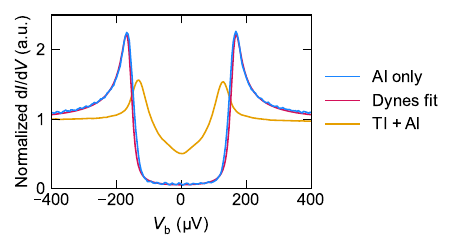}
\caption{\linespread{1.05}\selectfont{}
$\mathrm{d}I/\mathrm{d}V$  spectra measured at a different device with no TI coupled to the aluminum (blue curve) in comparison to the device studied here (yellow curve). The Al only data is well described by a standard Dynes model using $T_{\rm c}=\SI{1.05}{K}$, $T=\SI{0.05}{K}$, and $\Gamma=\SI{8.8}{\mu eV}$ }
\label{fig:SIgap}
\end{figure}

In Fig. \ref{fig:SIgap}, we show data from a different device, where the tunnel contact was not in contact with a topological insulator, but only probing the superconducting film itself. The resulting experimental curve is well described using the typical Dynes equation for the density of states $N_{\rm s}$ \cite{Dynes1978} and calculating the tunneling current by convolving it with the derivative of the Fermi-Dirac distribution at the given temperature.
\begin{align}
    N_{\rm s}(E)&\propto\left | \Re \left ( { \frac{E-i\Gamma}{\sqrt{(E-i\Gamma)^2-\Delta^2}}} \right ) \right | \\
\frac{dI}{dV}(V) &\propto \int_{-\infty}^{\infty} N_{\rm s}(E)\, \left(-\frac{\partial f(E - eV)}{\partial E}\right)\, dE
\end{align}
We find good agreement for $T_{\rm c}=\SI{1.05}{K}$, $T=\SI{0.05}{K}$, and $\Gamma=\SI{8.8}{\mu eV}$.  For comparison, the zero-field data of the tunneling conductance from our topological insulator device is also shown. 

\section{Sample geometry and flux focusing}
\label{SI:geom}
The inner area of the loop is given by $W_{\rm L} \times H_{\rm L} = 2.15 \times 3$ $\upmu$m$^2$, while the junction has a width of 3 $\upmu$m and a length of about 50 nm. The superconducting (SC) electrodes have a width of 400 nm. Making a simple assumption that the flux expulsion from the SC electrodes leads to an increase of the magnetic fluxes for which half the electrode width should be taken into account\cite{Ghatak2018}, we obtain an expected periodicity of  $\Delta B_1$ = 196 $\upmu$T for TJ1 and $\Delta B_2$ = 225 $\upmu$T for TJ2, in excellent agreement with our experimental data.

\section{Data in higher magnetic fields}
\label{SI:HighB}
\begin{figure}
\centering
\includegraphics{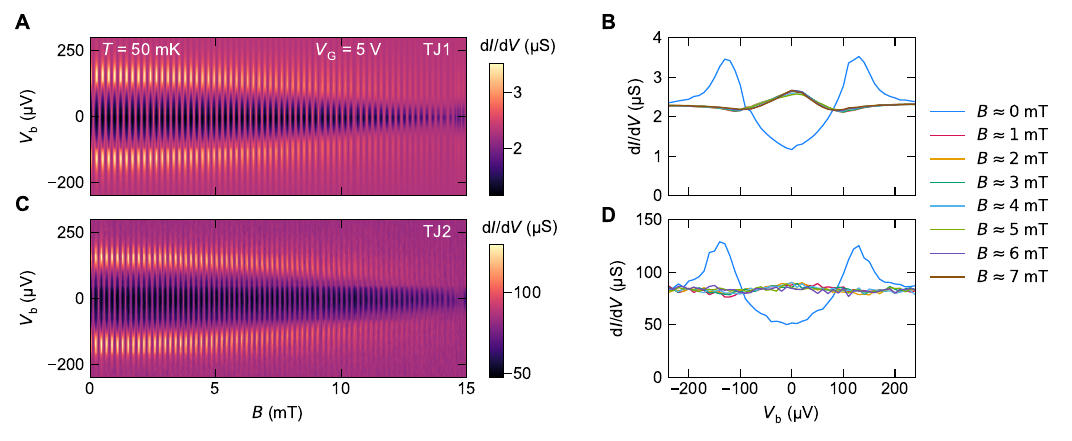}
\caption{\linespread{1.05}\selectfont{}
\textbf{A,C,} $\mathrm{d}I/\mathrm{d}V$  spectra measured at TJ1 and TJ2 of the device discussed in the main text (device 1) for the full magnetic-field range that was available in this experiment. A gradual reduction of the induced gap is observed in both junctions. \textbf{B,D,} Line cuts of the data in panels \textbf{A} and \textbf{B} at 0 T and at the fields near 1, 2, 3, 4, 5, 6, and 7 mT corresponding to $\varphi\approx\pi$.}
\label{fig:SI2}
\end{figure}

In Fig. \ref{fig:SI2}, we show the $\mathrm{d}I/\mathrm{d}V$  spectra on TJ1 and TJ2 from 0 T to 15 mT, which was the maximum available magnetic field in this experiment due to the limitation of the high-resolution current source (Keithley 2450) used for energizing the superconducting magnet in the very low magnetic-field range. The superconductivity in the Al electrode was strongly suppressed in 15 mT, which can be seen in the suppression of the induced gap. Nevertheless, the dispersion of modes with $\tau>0.95$ can be observed in the whole magnetic-field range. For our Al films of thickness 45 nm, we typically observe a critical field of $\sim$20 mT. In the right panels we show line cuts at 0 T and at the fields corresponding to $\varphi\approx\pi$ near 1, 2, 3, 4, 5, 6, and 7 mT to demonstrate the robustness with respect to magnetic fields.

\section{Data for higher bias voltages}
\label{SI:HighVB}
\begin{figure}
\centering
\includegraphics{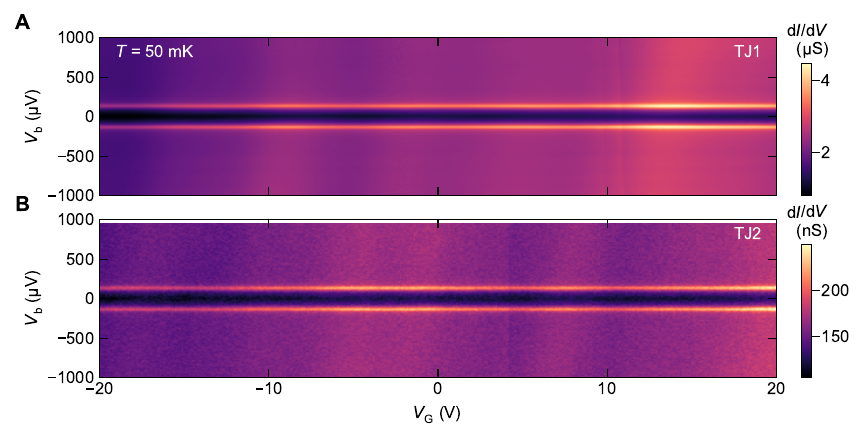}
\caption{\linespread{1.05}\selectfont{}
$\mathrm{d}I/\mathrm{d}V$  spectra measured at TJ1 (\textbf{A}) and TJ2 (\textbf{B}) of the device discussed in the main text (device 1) for a larger bias-voltage range ($\pm$1 mV) and without normalizing the data. }
\label{fig:SIHB}
\end{figure}

In Fig. \ref{fig:SIHB}, we show the $\mathrm{d}I/\mathrm{d}V$  spectra on TJ1 and TJ2 for the biases voltage $V_{\rm b}$ up to 1 mV for the entire range of $V_{\rm G}$. The general reduction of $\mathrm{d}I/\mathrm{d}V$  at more negative $V_{\rm G}$ values, reflecting the $n$-type nature of the TI flake used here, is clearly visible. Also, one can see a sequence of broad impurity resonances occasionally crossing the gap energy for both devices without introducing in-gap states. In both cases, it requires $\Delta V_{\rm G} \approx1-2\,\mathrm{V}$ to change their energy by $2\,\mathrm{meV}$, resulting in the gate efficiency given in the main text.

\section{Josephson penetration length}
\label{SI:JoPen}
The RF-SQUID geometry of our device does not allow for a direct measurement of its critical current. However, in our previous study of similar devices \cite{Ghatak2018}, we typically find a critical current per unit width of the Josephson junction $I_{\rm c}/W\approx 100\, \mathrm{nA/ \upmu m}$. We use this value to estimate the Josephson penetration length in our device using $\lambda_{\rm J}=\sqrt{\frac{\Phi_0W^2t}{4\pi\mu_0 I_{\rm c}  \lambda^2}}$, where $W=3\,\mathrm{\upmu m}$ is the width of the junction, $t=2\,\mathrm{nm}$ is the extension of the surface state \cite{Zhang2010}, and $\lambda\approx100\,\mathrm{nm}$ is the London penetration depth for an aluminum film of $\sim$50-nm thickness \cite{Lopez2023}. We obtain $\lambda_{\rm J } \approx 28\,\mathrm{\upmu m}$, approximately one order of magnitude larger than our junction width, justifying the assumption of the linear phase gradient in the junction for finite magnetic fields.

\section{Loop inductance}
\label{SI:LI}
The geometrical inductance of our RF-SQUID can be estimated using the model cited e.g. in Ref. \cite{Shatz2014}. It yields $L_{geo}\approx\SI{6}{pH}$ for the dimensions of our devices. The kinetic inductance of Al films of comparable fabrication and thickness was determined in Ref. \cite{Lopez2023}. Using their values, we estimate $L_{kin}\approx \SI{9}{pH}$ for our device.
Even if we assume a large critical current of $I_{\rm c}\approx\SI{1}{\mu A}$, the ratio $(L_{geo}+L_{kin})I_{\rm c}/\Phi_0\approx 0.008<<1$ indicates that the loop inductance will have a marginal impact.

\section{Surface state coherence length}
\label{SI:sscl}
The Fu-Kane theory  \cite{Fu2008} assumed that the Josephson junction is in the short-junction limit. This requires the superconducting coherence length $\xi_{\rm TI}$ in the TI surface state to be larger than the junction length $L=50\,\mathrm{nm}$. The effective coherence length in a non-ballistic junction can be calculated via $\xi_{\rm TI} = \sqrt{\frac{\hbar v_{\rm F} l_{\rm e}}{2\Delta}}$, with $l_{\rm e}$ the mean free path in the TI surface. 
If we take $l_{\rm e} \approx$ 40 nm and $v_{\rm F} = 4.6 \times 10^5$ m/s from Ref.  \citenum{Taskin2011}, we obtain $\xi_{\rm TI} \approx$ 200 nm, which is longer than the junction length.   
Even if we take a much shorter $l_{\rm e}\approx 5\,\mathrm{nm}$ combined with $v_{\rm F} = 5.5 \times 10^5$ m/s as in Ref.  \citenum{Ghatak2018}, we obtain $\xi_{\rm TI} \approx 75\,\mathrm{nm}$, which is still longer than the junction length.   
Therefore, our junctions are safely in the short-junction limit.

\section{Temperature dependence of the spectra}
\label{SI:Tdep}
\begin{figure}
\centering
\includegraphics[width=\textwidth]{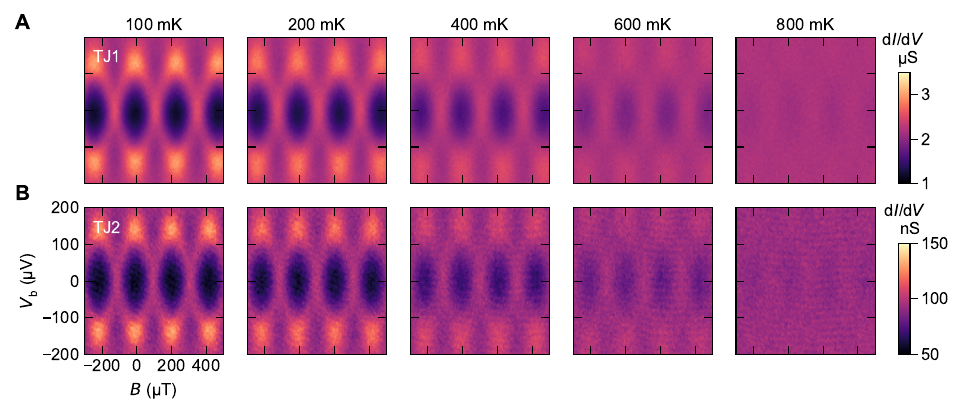}
\caption{\linespread{1.05}\selectfont{}
Evolution of the spectra for TJ1 (\textbf{A}) and TJ2 (\textbf{B}) at $V_{\rm G}=0$ with increasing temperature. }
\label{fig:SI3}
\end{figure}

In Fig. \ref{fig:SI3}, we show the temperature dependence of the spectra for TJ1 and TJ2 at $V_{\rm G}=0$. The spectra are essentially robust against increasing temperature, although thermal smearing is evident. Even at $T=600\,\mathrm{mK}$, the key features are still recognizable. 

\section{Confirmation of reproducibility}
\label{SI:repro}
\begin{figure}
\centering
\includegraphics{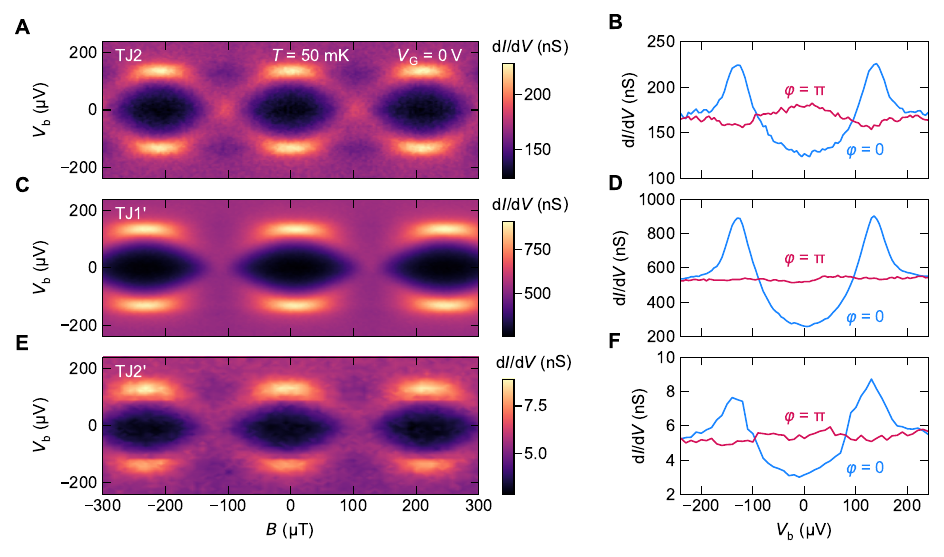}
\caption{\linespread{1.05}\selectfont{}
\textbf{A, B,}Tunnel spectra for TJ2 of the device discussed in the main text. \textbf{C-F,} Data on TJ1' and TJ2' made on an identically-designed device fabricated on a different flake of the same crystal. $V_{\rm G}$ = 0 V for all the data.  In panels  \textbf{B},  \textbf{D}, and  \textbf{F}, line cuts at $\varphi=0$ (blue) and $\varphi=\pi$ (red) are shown for each junction. }
\label{fig:SI1}
\end{figure}

In Fig. \ref{fig:SI1}, we show the data for TJ2 on the device discussed in the main text (device 1) at $V_{\rm G}=0$, as well as the data obtained from another device with an identical design (device 2). It is fabricated on a BiSbTeSe$_2$ flake exfoliated from the same crystal and has tunnel junctions TJ1' and TJ2' in analogous positions. We observe the same closing and reopening of the gap as a function of $\varphi$ as for device 1.

\begin{figure}
\centering
\includegraphics{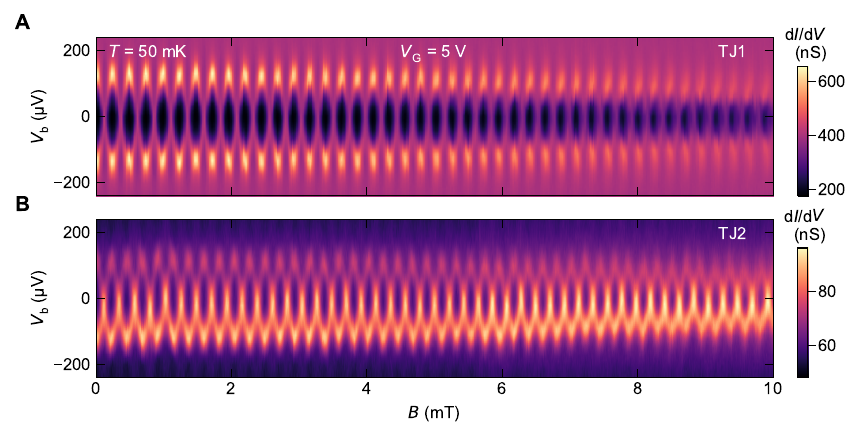}
\caption{\linespread{1.05}\selectfont{}
 $\mathrm{d}I/\mathrm{d}V$  spectra measured at TJ1' (\textbf{A}) and TJ2' (\textbf{B}) for a higher magnetic field range, analogue to the data shown in Fig. \ref{fig:SI2} for device 1. The data were taken with a different current preamplifier, showing a much stronger time-dependent drift of the offset voltage that had to be numerically corrected. }
\label{fig:SIffdev2}
\end{figure}

Similarly, the locality of the spectra is also reproduced as shown in Fig. \ref{fig:SIffdev2}. The periodicity for TJ1' is larger than that for TJ2', which is understood from the geometry as discussed in the main text.

\begin{figure}
\centering
\includegraphics{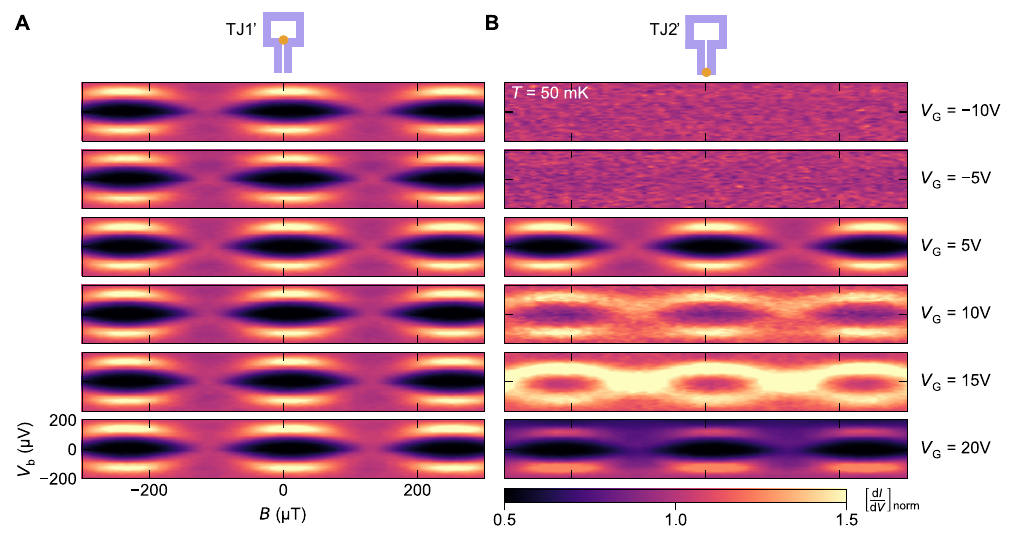}
\caption{\linespread{1.05}\selectfont{}
Spectra of TJ1' (\textbf{A}) and TJ2' (\textbf{B}) of device 2. For negative gate voltages, TJ2' becomes unmeasurable due to its conductance dropping below the noise floor. }
\label{fig:SIgatedev2}
\end{figure}

In Fig.~\ref{fig:SIgatedev2}, the gate stability of the data from device 2 is shown. For TJ1', the phenomenology is the same as discussed for device 1. We observe an apparent gap closing at $\varphi = (2n+1)\pi$, which in some cases (e.g. $V_{\rm G}$ = 5 V) is accompanied by a broad ZBCP. For TJ2', the situation is more complex. Its conductance becomes strongly suppressed for negative gate voltages, and the strongly asymmetric signals for more positive gate voltages suggests a highly disordered dielectric for this tunnel contact. Nevertheless, the apparent gap closing itself is reproduced.

\section{Additional device with extended gate range}
\label{SI:VgX}
\begin{figure}
\centering
\includegraphics{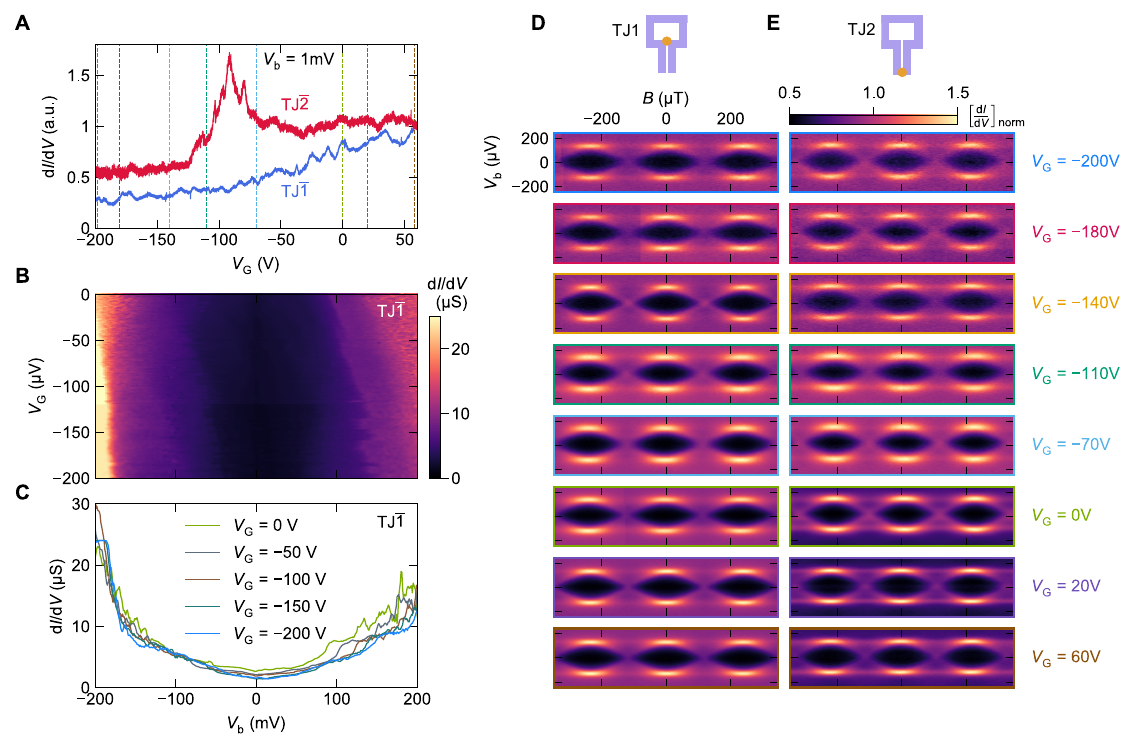}
\caption{\linespread{1.05}\selectfont{}
{\bf Device with extended gate range. } 
\textbf{A,} Above gap differential conductance of TJ$\overline{1}$ and TJ$\overline{2}$, normalized to their respective values at 60\,V. \textbf{B,} Colorplot of the differential conductance of TJ$\overline{1}$ measured over a range of bias voltages that cover the bulk band gap of the topological insulator. \textbf{C,} Line cuts of \textbf{B}, showing the systematic shift of the feature we interpret as bottom of the conduction band. \textbf{D,E,} Normalized differential conductance for TJ$\overline{1}$ and TJ$\overline{2}$, respectively, for a selection of gate voltages.  }
\label{fig:SI18}
\end{figure}

The risk of electrostatic discharge as well as potentially leaky gate dielectrics limited the gating range of the device discussed in the main text.
In an additional device fabricated in the same way, we were able to tune the gate voltage over a wider range. In Fig. \ref{fig:SI18}a, we show the normalized differential conductance for both tunnel junctions on this device for -200\,${V}\leq V_{\rm G}\leq$ 60\,$V$. For more negative gate voltages, the gate dielectric started to develop notable leakage current that impeded further measurement. We observe a decrease of conductance by a factor of 2 and 4 at the two tunnel electrodes, larger than the one presented in the main text. However, contrary to the situation in Ref. \cite{Ghatak2018}, a global minimum, indicating the Dirac point, is not found. We tentatively attribute this to the reduced gating efficiency due to the screening from the metallic tunneling electrode.

To obtain more quantitative information on the position of the chemical potential with respect to the bulk band structure, we performed extended bias sweeps, shown in Figs. \ref{fig:SI18}b,c. The bias range of $\pm$200\,mV studied here should well cover the bulk band gap of $\text{BSTS}_2$, which was shown to be about 300\,meV large \cite{Arakane2012}. The overall gate-tunability is well visible. In particular, the sudden increase in conductance at 90\,meV for $V_{\rm G}=$0\,V clearly shifts towards higher energies with more negative gate voltages. One can reasonably assign this feature to the bottom of the conduction band \cite{Brede2024}. At $V_{\rm G}=$-200\,V, this feature appears at $V_{\rm b}=$150\,mV, indicating a shift in the chemical potential of $\Delta\mu=$60\,meV in the range of $\Delta V_{\rm G}=$200\,V. From this we can infer $\mu<$50\,meV, following the results of Ref. \cite{Arakane2012}. A clearer observation of the bulk band structure is probably impeded by the non-ideal tunnel contact as well as the presence of the Al-electrodes nearby.\\
Reassuringly, the appearance of strongly dispersing modes, potentially indicating a gap closing and reopening phenomenology, shown in Fig. \ref{fig:SI18}D,E, remains robust also over the entire range of chemical potential studied in this device. 

\section{Observation of spurious resonant states}
\label{SI:QDs}
\begin{figure}
\centering
\includegraphics{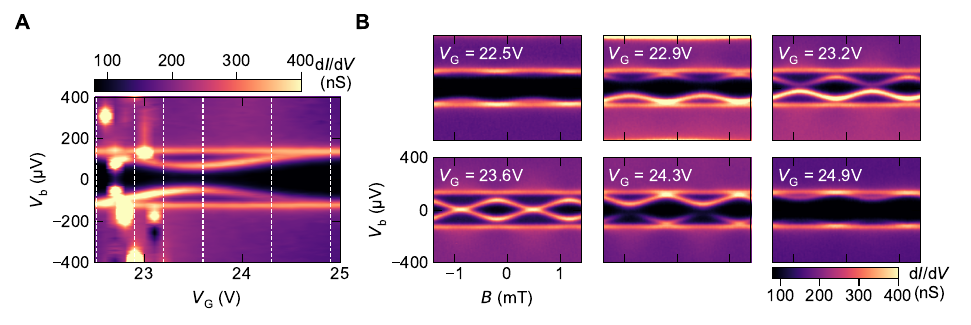}
\caption{\linespread{1.05}\selectfont{}
{\bf Spurious resonance state. } 
\textbf{A,} $\mathrm{d}I/\mathrm{d}V$ spectra as a function of $V_{\rm G}$ and $V_{\rm b}$ measured at $\varphi$ = 0 in a device similar to the one discussed in the main text. In a narrow $V_{\rm G}$ range of 22.5 -- 25 V, a single ABS appeared in the SC gap.  \textbf{B,} $\varphi$-dependence of the ABS spectra at selected $V_{\rm G}$ values in this narrow $V_{\rm G}$ range. }
\label{fig:SI4}
\end{figure}

In a device of similar geometry, we observed an induced SC gap, but the spectrum did not change with the phase bias at $V_{\rm G} = 0$. However, when we swept $V_{\rm G}$ without applying any phase bias, a sharp and strongly-dispersing Andreev bound state (ABS) appeared in some $V_{\rm G}$ intervals. An example is shown in Fig. \ref{fig:SI4}a for the $V_{\rm G}$-interval of 22.5 -- 25 V. 
The phase-bias dependence of this ABS changes strongly with $V_{\rm G}$ (Fig. \ref{fig:SI4}b), and interestingly, the state becomes gapless at $V_{\rm G}$ = 23.6 V where it produced a sharp ZBCP. The electron-hole asymmetry of the ABS switched across $V_{\rm G}$ = 23.6 V. This behavior is consistent with the model for a quantum-dot state in a Josephson junction being tuned in and out of resonance with the superconductor  \cite{Oriekhov2021}. It is useful to note that this ZBCP appears only at particular $V_{\rm G}$ values and is not accompanied by low-energy excited states, which is completely different from the gap closing phenomenology expected in the Fu-Kane model.  
As for the origin of this peculiar feature observed in this device, we speculate that due to the uncontrollable nature of our tunnel contacts, the tunnel electrode spuriously did not couple to the TI surface, but rather coupled to a single impurity level in the dielectric with a charging energy that was tuned by $V_{\rm G}$.

\begin{figure}
\centering
\includegraphics{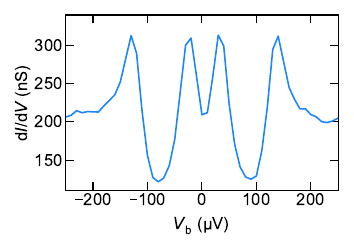}
\caption{\linespread{1.05}\selectfont{}
{\bf Energy resolution } 
Line-cut of the spectrum of the Andreev quantum dot at resonance $V_{\rm G}=23.6\,\mathrm{V}$ close to (but not exactly at) $\varphi=\pi$. The individual peaks located at $V_{\rm b} \approx \pm 30\, \upmu \mathrm{eV}$ are clearly resolved. }
\label{fig:SIReso}
\end{figure}

\begin{figure}
\centering
\includegraphics{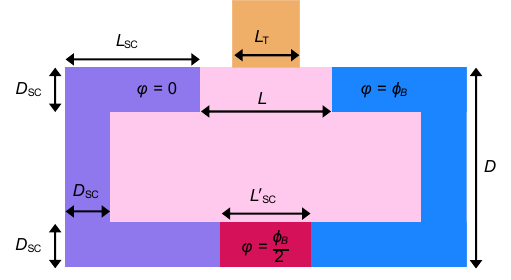}
\caption{\linespread{1.05}\selectfont{}
{\bf Sketch of the simulated system.} The cross section of the simulated system shown with its dimensions. The bulk TI in the center (light magenta-tinted pink) is surrounded by three separate regions with finite superconducting gap $\Delta$ (lavender, magenta, azure blue). The phase of the gap function $\varphi$ differs between the regions. On the top surface, this creates a Josephson junction with a finite width $W$ and the phase difference $\phi_B$. A tunnel lead (light orange) is attached in the center of the bare TI surface of the junction.}
\label{fig:SI5}
\end{figure}

We can make use of this spurious feature to estimate the energy resolution. In Fig. \ref{fig:SIReso} we present a single line cut at resonance $V_{\rm G}$ = 23.6 V, close to $\varphi=\pi$ but slightly off. In this plot, the two ABSs are individually resolved and the peaks appear at $\pm 30\,\upmu\mathrm{eV}$. We therefore conclude that our energy resolution is of the order of  $30\,\upmu\mathrm{eV}$, which is much larger than the thermal energy of the measurement temperature ($\sim$50 mK) and we therefore attribute it to electronic noise in the setup.

\section{Calculation of the Majorana velocity}
\label{SI:vm}
The Majorana velocity for the Josephson junction in the short-junction limit depends on the length of the junction.
Starting from the surface Hamiltonian of the TI and including superconductivity in the Bogoliubov-de Gennes formalism, the full Hamiltonian can be split into
\begin{equation}
    H = \underbrace{H_\perp + H_\pi}_{H_\text{JR}} + H_\parallel  \, ;
\end{equation}
$H_\perp$ and $H_\parallel$ denote the momentum component orthogonal and parallel to the junction, respectively, and $H_\pi$ accounts for the superconductivity for a phase difference of $\varphi=\pi$ across the junction. 
$H_\text{JR}=H_\perp + H_\pi$ is a four-band generalization of the Jackiw-Rebbi model and can be solved exactly for two zero-energy Majorana bound states in the junction.
The remaining contributions are included perturbatively and by evaluating $H_\parallel$ to first order we find the Majorana velocity
\begin{equation}
    v_{\rm M} = v_{\rm F} \frac{\Delta^2}{\Delta^2+\mu^2} \, \frac{\cos\bigl(\tfrac{\mu L}{\hbar v_{\rm F}}\bigr)+\frac{\Delta}{\mu} \sin\bigl(\tfrac{\mu L}{\hbar v_{\rm F}}\bigr)}{1+\tfrac{\Delta L}{\hbar v_{\rm F}}}
\end{equation}
with the TI surface velocity $v_{\rm F}$, the magnitude of the superconducting gap $\Delta$, the chemical potential $\mu$, and the length of the junction $L$. For more information, see Eq.~(3.23) of Ref.~ \citenum{AntonThesis}. In the limit $L \ll \hbar v_{\rm F}/\Delta$, it reduces  to the result presented by Fu and Kane in Ref.~ \citenum{Fu2008}. Moreover, for $\mu \gg \Delta$, we obtain the simple expression $v_{\rm M} = v_{\rm F} (\Delta/\mu)^2$ mentioned in Ref.  \citenum{Potter2013}. We note that the assumption of a single transversal mode, which lies behind the calculations, is valid for $\mu < \hbar v_{\rm F}/L $.

\section{Theoretical simulation at fixed phase}
\label{SI:simfix}
The theoretical simulations presented in the main text were performed by taking a running average over the phase in a small window of $\delta \varphi \approx 0.15 $.
This approach was adopted to capture the phase uncertainty (either along the junction or in time) of the experimental situation.
The peak at zero bias seen in Fig.~2(b) for $\varphi=\pi$ in the main text is not a signature of a zero mode.
Rather, it is a broadened version of a BCS peak, which appears for any slight deviation from the exact value of $\varphi=\pi$.

\begin{figure}
\centering
\includegraphics{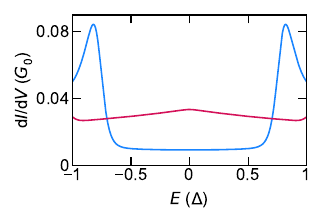}
\caption{\linespread{1.05}\selectfont{}
{\bf Theoretical simulation without phase averaging.} This plot corresponds to Fig. 2(b) in the main text. The spectra for $\varphi=0$ (blue) and $\varphi=\pi$ (red) were calculated without applying the running phase average.  The blue line remains essentially unchanged, while the red line no longer shows a zero-bias peak.}
\label{fig:SI6}
\end{figure}

To demonstrate this point, we show in Fig.~\ref{fig:SI6} the result of the same simulation as that for Fig.~2(b), but without applying a running average over the phase.
Note that the curve for $\varphi=0$ (blue) remains essentially unchanged, whereas the zero-bias peak for $\varphi=\pi$ disappears in Fig.~\ref{fig:SI6} and is replaced by a nearly flat dependence (red).
This calculation clearly shows that the zero-bias feature at  $\varphi=\pi$ is highly sensitive to experimental conditions, such as whether there is an exact phase bias or some averaging over a small phase window.
Importantly, this sensitivity of the spectrum to a small phase variation occurs only near $\varphi=\pi$, while the general features of the gap closing discussed in the main text remain robust.  
\newpage

\end{document}